\documentclass[review]{elsarticle}

\AtBeginDocument{}

\usepackage{textcomp}
\usepackage{graphicx}
\usepackage{enumitem}
\usepackage{longtable}
\usepackage{subcaption}
\usepackage{alphalph}
\usepackage[many]{tcolorbox}
\usepackage{xcolor}
\biboptions{sort&compress}
\usepackage{array}
\usepackage{colortbl}

\newcolumntype{L}{>{\arraybackslash}m{16cm}}
\usepackage[flushleft]{threeparttable}
\newcolumntype{C}[1]{>{\centering\let\newline\\arraybackslash\hspace{0pt}}m{#1}}
\newcolumntype{R}[1]{>{\raggedleft\let\newline\\arraybackslash\hspace{0pt}}m{#1}}
\def\BibTeX{{\rm B\kern-.05em{\sc i\kern-.025em b}\kern-.08em
    T\kern-.1667em\lower.7ex\hbox{E}\kern-.125emX}}
\usepackage{csquotes}

\usepackage{tikz}
\usetikzlibrary{fit} 
\usetikzlibrary{positioning}
\usetikzlibrary{arrows}
\usetikzlibrary{shapes.multipart}

\usepackage[hidelinks,bookmarks=false]{hyperref}
\usepackage[numbered]{bookmark}
\usepackage{soul}
\usepackage{booktabs}
\usepackage{multirow}
\usepackage{float}
\usepackage{url}
\usepackage{amsmath,amssymb,amsfonts}
\usepackage{algorithmic}
\usepackage{textcomp}
\usepackage{url}
\usepackage{longtable}
\journal{Journal of \LaTeX\ Templates}

\usepackage{pgfplots}
\pgfplotsset{width=7cm,compat=1.8,tick label style={font=\small}}

\usepackage{xspace}
\newcommand{\ie}{\textit{i.e., \xspace}}
\newcommand{\eg}{\textit{e.g., \xspace}}
\newcommand{\etal}{\textit{et al. \xspace}}
\usepackage{adjustbox}
\usepackage{float}
\usepackage{rotating}
\usepackage{tablefootnote}
\usepackage{nth}
\DeclareCaptionType{TextBox}

\usepackage[utf8]{inputenc}
\usepackage{dirtytalk}
\usepackage{tcolorbox}

\usepackage[british]{babel}
\usepackage{enumitem}

\newlist{SubItemList}{itemize}{1}
\setlist[SubItemList]{label={$-$}}

\let\OldItem\item
\newcommand{\SubItemStart}[1]{%
    \let\item\SubItemEnd
    \begin{SubItemList}[resume]%
        \OldItem #1%
}
\newcommand{\SubItemMiddle}[1]{%
    \OldItem #1%
}
\newcommand{\SubItemEnd}[1]{%
    \end{SubItemList}%
    \let\item\OldItem
    \item #1%
}
\newcommand*{\SubItem}[1]{%
    \let\SubItem\SubItemMiddle%
    \SubItemStart{#1}%
}

\newtcolorbox{boxK}{
    sharpish corners, 
    boxrule = 0pt,
    toprule = 4.5pt, 
    enhanced,
    fuzzy shadow = {0pt}{-2pt}{-0.5pt}{0.5pt}{black!35} 
}

    \newcommand{\RQone}{How do refactoring reviews compare to non-refactoring reviews
in terms of code review efforts?\xspace}

\newcommand{\RQthree}{What quality attributes do developers consider
when describing refactoring in the `Refactor’ branch?\xspace}

\newcommand{\RQtwo}{What textual patterns do developers use to describe their refactoring needs in the `Refactor’ branch?\xspace}

\newcommand{\RQfour}{What topics do developers discuss when reviewing refactoring tasks?\xspace}

\begin{document}





\begin{frontmatter}

\title{Deciphering Refactoring Branch Dynamics in Modern Code Review: An Empirical Study on Qt}

\author[RIT]{Eman Abdullah AlOmar\corref{mycorrespondingauthor}}
\cortext[mycorrespondingauthor]{Corresponding author}
\ead{ealomar@stevens.edu}

\address[RIT]{Stevens Institute of Technology, Hoboken, NJ, USA}

\begin{abstract}
\noindent\textbf{Context:} Modern code review is a widely employed technique in both industrial and open-source projects, serving to enhance software quality, share knowledge, and ensure compliance with coding standards and guidelines. While code review is extensively studied for its general challenges, best practices, outcomes, and socio-technical aspects, little attention has been paid to how refactoring is reviewed and what developers prioritize when reviewing refactored code in the `Refactor' branch. 

\noindent\textbf{Objective:} \textcolor{black}{The goal is to understand the review process for refactoring changes in the ‘Refactor’ branch and to identify what developers care about when reviewing code in this branch.}

\noindent\textbf{Method:} In this study, we present a quantitative and qualitative examination to understand the main criteria developers use to decide whether to accept or reject refactored code submissions and identify the challenges inherent in this process.

\noindent\textbf{Results:} Analyzing 2,154 refactoring and non-refactoring reviews across Qt open-source projects, we find that reviews involving refactoring from the `Refactor' branch take significantly less time to resolve in terms of code review efforts. Additionally, documentation of developer intent is notably sparse within the `Refactor' branch compared to other branches. Furthermore, through thematic analysis of a substantial sample of refactoring code review discussions, we construct a comprehensive taxonomy consisting of 12 refactoring review criteria.

\noindent\textbf{Conclusion:} Our findings underscore the importance of developing precise and efficient tools and techniques to aid developers in the review process amidst refactorings.
\end{abstract}


\end{frontmatter}


\section{Introduction}
\label{Section:Introduction}


Refactoring is a crucial practice for maintaining code quality as software evolves. Its significance has expanded beyond mere code cleanup to become a cornerstone of modern software development. This has attracted significant attention from researchers, evident in the numerous research papers dedicated to the topic \cite{abid202030}. Another key practice in maintaining software quality is code review \cite{bacchelli2013expectations}. 
 It has become another important to reduce technical debt, and to detect potential coding errors \cite{bacchelli2013expectations,sadowski2018modern,kashiwa2022empirical}. Code review represents the manual inspection of any newly performed changes to the code, for the purpose of verifying integrity, compliance with standards, and error-freedom \cite{mcintosh2016empirical}. Today's \textcolor{black}{Modern Code Review (MCR)} process is typically lightweight and tool-based, relying heavily on discussions between authors and reviewers to decide whether to merge or discard a code change \cite{yang2016mining}.

Refactoring changes, like any other code modifications, must undergo review before being merged. Failure to apply refactoring properly can lead to adverse effects, including compromised software quality \cite{hamdi2021empirical,alomar2019impact,hamdi2021longitudinal,peruma2020exploratory} and inducing bugs \cite{bavota2012does,di2020relationship} 
 making refactoring changes more challenging to review. However, little is known about how reviewers \textit{examine} refactoring related code changes, especially when it is intended to serve the same \textit{purpose} of improving software quality. According to the industrial case study, AlOmar \etal \cite{alomar2021icse} has found that reviewing refactoring related code changes takes a significantly longer time, in comparison with other code changes, demonstrating the need for refactoring review \textit{culture}. 
  Yet, little is known about what criteria reviewers consider when they review refactoring. Most of refactoring studies focus on its automation by recommending refactoring opportunities in the source code \cite{tsantalis2008jdeodorant,mkaouer2015many,ouni2016multi}, or mining performed refactorings in change histories of software repositories \cite{tsantalis2018accurate}. 
  Moreover, while research on code reviews has concentrated on automation, such as recommending the most suitable reviewer for a given code change \cite{bacchelli2013expectations}, the review process for refactoring changes in the `Refactor' branch remains largely unexplored. 

\begin{figure*}[th]
\centering 
\includegraphics[width=\textwidth]{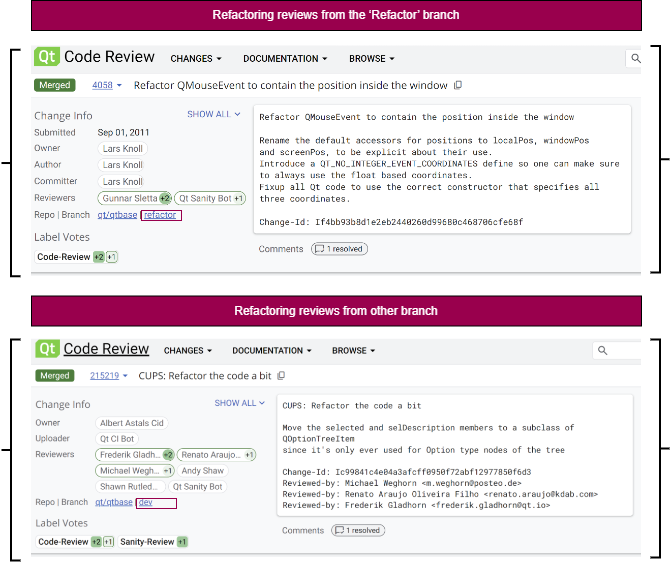}
\vspace{-.5cm}
\caption{\textcolor{black}{Example of a code review from Qt project using Gerrit \cite{Example,Example2}.}}
\label{fig:example}
\vspace{-.4cm}
\end{figure*}

Building upon our previous research \cite{alomar2021icse,alomar2022code}, which revealed that refactoring code reviews often take longer to be approved, we now delve deeper into the specific practices of refactoring review within an ecosystem featuring a dedicated `Refactor' branch. This investigation aims to provide insights into addressing the challenges identified in our previous studies, thus contributing to a more efficient and effective refactoring review process.  \textcolor{black}{Specifically, the goal of this paper is to understand the criteria developers use when reviewing refactored code in the `Refactor' branch, focusing on what influences their decisions to accept or reject submissions.}

 \textcolor{black}{In our study, we explore two distinct venues where refactoring can occur: the typical branch and the refactoring branch. The typical branch refers to the standard development workflow where incremental changes and features are integrated. In contrast, the refactoring branch is specifically dedicated to restructuring and improving the codebase without altering its external behavior. Figure \ref{fig:example} illustrates a segment of the refactoring review from the `Refactor' branch and other branches within the Qt system. The top example shows a code review that was created to refactor \texttt{QMouseEvent} to contain the position inside the window. The developers submitted the code, while explicitly stating that the refactoring was intended to rename the default accessors for positions to \texttt{localPos}, \texttt{windowPos} and \texttt{screenPos}, to be explicit about their use. This can also be seen in the final subject and description of the review that later merged the modified code into production. Based on this example, we can see that \textit{rename} was one of the refactoring operations that developers consider for code optimization, and author confirms its quality improvement as follows \say{\textit{There's no behavioural change. All I did was add the windowPos(), and make sure it's set correctly everywhere. Old code will continue to work just as before.}} In contrast, the bottom example reveals changes to the \texttt{QtBase} module. It focuses on enhancing the code by removing unnecessary elements, improving readability, and making the code more maintainable. The review includes detailed comments from developers discussing the impact and necessity of these changes, aiming to ensure that the refactoring maintains the module's functionality while improving its structure. This review had a longer review duration and more extensive discussion compared to the `Refactor' branch reviews. We believe that early clarification of the `Refactor' and the typical branch is crucial for understanding their roles in the development process. The typical branch often prioritizes feature development and bug fixes, while the refactoring branch focuses on enhancing code quality, maintainability, and performance. By distinguishing between these two branches, we aim to highlight the unique challenges and considerations associated with each, thereby providing a comprehensive understanding of their impact on the software development lifecycle.} 

 \textcolor{black}{Therefore, we conduct our study using the following 
 overarching question: \textit{How do developers approach and evaluate refactoring tasks compared to non-refactoring tasks during code reviews, and what patterns, quality attributes, and topics are emphasized in these reviews?}}

To answer our research questions, we first extracted a set of 718 refactoring-related code reviews in the `Refactor' branch from the Qt ecosystem. 
 Then, we compared this set of refactoring-related code reviews in `Refactor' branch, with another two sets of code reviews, in terms of the number of reviewers, number of review comments, number of inline comments, number of revision, number of changed files, review duration, discussion and description length, and code churn. 
  Our empirical investigation indicates that refactoring-related code reviews from `Refactor' branch
take significantly shorter to be resolved and typically trigger fewer discussions between developers and reviewers to reach a consensus. To understand the key characteristics of reviewing refactored code, we perform a thematic analysis on a significant sample of these reviews. This process resulted in a hierarchical taxonomy composed of four categories, and 12 sub-categories. 

We provide our experiments package \cite{ReplicationPackage} to further replicate and extend our study. The package contains raw data, analyzed data, statistical test results, survey questions, and custom-built scripts used in our research.

The remainder of this paper is organized as follows. Section \ref{Section:Background} provides background on the Gerrit-based code review process. Section \ref{Section:RelatedWork} reviews the existing studies related to refactoring awareness and code review. Section \ref{Section:Methodology} outlines our empirical setup in terms of data collection, analysis and research question. Section \ref{Section:Result} discusses our findings, while the research implication is discussed in Section \ref{Section:Implications}. Section \ref{Section:Threats} captures any threats to the validity of our work, before concluding with Section \ref{Section:Conclusion}.
\section{\textcolor{black}{Background}}
\label{Section:Background}

 \begin{figure}[]
 	\centering
 	\includegraphics[width=1.0\columnwidth]{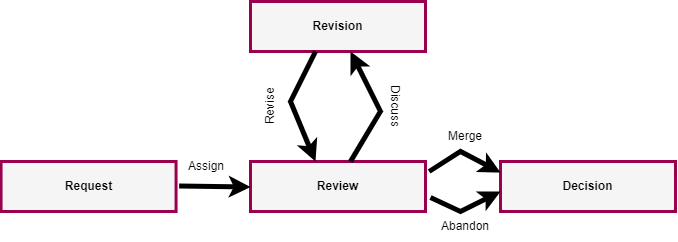}
 	\caption{\textcolor{black}{Gerrit-based code review process overview.}} 
 	\label{fig:review}
\vspace{-.4cm}
\end{figure}

\textcolor{black}{Code review involves the manual assessment of source code by humans to identify defects and quality issues \cite{beller2014modern}. However, traditional code review practices have limitations when applied to globally distributed software development \cite{votta1993does}. In recent years, Modern Code Review (MCR) has emerged as a tool-based system that is less formal than traditional methods. MCR has become popular and is widely used in both proprietary software (\eg Google, Microsoft) and open-source software (\eg OpenStack, Qt) \cite{bacchelli2013expectations}. In this study, we selected Gerrit as it provides robust code review functionality that is essential to maintain code quality and ensure thorough review processes. This feature is crucial for our research's needs, where code integrity and peer review are paramount. In the following, we provide a brief overview of the Gerrit-based code review system, a prominent tool frequently used in previous studies \cite{rigby2013convergent,mcintosh2014impact,thongtanunam2014reda,kula2012using}.}

The code review process of the systems studied is based on Gerrit\footnote{\url{https://www.gerritcodereview.com/}}, a collaborative code review framework.   Gerrit facilitates developers in tagging submitted code changes directly and requesting their assignment to a reviewer. Generally, a code change author opens a code review request containing a title, a detailed description of the code change being submitted, written in natural language, and the current code changes annotated. Once the review request is submitted, it will appear in the requests backlog and be open to reviewers to choose from. Once reviewers are assigned to the review request,
they inspect the proposed changes and comment on the review request's thread to start a discussion with the author. This way, the authors and reviewers can discuss the submitted changes, and reviewers can request revisions to the code being reviewed. Following up discussions and revisions, a review decision is made to either accept or decline, and so the proposed code changes are either \say{\textit{Merged}} to production or \say{\textit{Abandoned}}. 

\textcolor{black}{A diagram, modeling a simplified bird's view of the Gerrit-based code review process, is shown in Figure \ref{fig:review}. It begins with a \say{Request} for a review, which is then assigned to a reviewer. The \say{Review"} phase follows, where the code is assessed and discussed. If changes are needed, the code enters the \say{Revision} phase, where it is revised based on feedback and can be discussed further. This cycle continues until a decision is made to either \say{Merge} the code into the main branch or \say{Abandon} the changes. }

\section{\textcolor{black}{Related Work}}
\label{Section:RelatedWork}

Research on code review has been of importance to practitioners and researchers. A considerable effort has been spent by the research community in studying traditional and modern code review practices and challenges. This literature has included case studies  (\eg  \cite{ge2014towards,mcintosh2014impact,ge2017refactoring,morales2015code,sadowski2018modern,rigby2013convergent,alomar2021icse}), user studies (\eg \cite{barnett2015helping,tao2015partitioning,zhang2015interactive,alves2017refactoring,peruma2022refactor}), surveys (\eg \cite{tao2012software,bacchelli2013expectations,macleod2017code,alomar2021icse}), and empirical experiments (\eg \cite{mcintosh2014impact,morales2015code,tao2015partitioning,guo2017interactively,peruma2019contextualizing}). However, most of the above studies focus on studying and improving the effectiveness of modern code review in general, as opposed to our work that focuses on understanding developers' perception of code review involving refactoring. In this section, we are only interested in research related to refactoring-aware code review. 

In a study performed at Microsoft, Bacchelli and Bird \cite{bacchelli2013expectations} observed and surveyed developers to understand the challenges faced during code review. They pointed out purposes for code review (\eg improving team awareness and transferring knowledge among teams) along with the actual outcomes (\eg creating
awareness and gaining code understanding). In a similar context, MacLeod \etal  \cite{macleod2017code} interviewed several teams at Microsoft and conducted a survey to investigate the human and social factors that influence developers' experiences with code review. Both studies found the following general code reviewing challenges: (1) finding defects, (2) improving the code, and (3) increasing knowledge transfer. Ge \etal  \cite{ge2014towards,ge2017refactoring} developed a refactoring-aware code review tool called ReviewFactor that automatically detects refactoring edits and separates refactoring from non-refactoring changes with a focus on five refactoring types. The tool was intended to support developers' review process by distinguishing between refactoring and non-refactoring changes, but it does not provide any insights on the quality of the performed refactoring. Inspired by the work of \cite{ge2014towards,ge2017refactoring}, Alves \etal  \cite{alves2014refdistiller,alves2017refactoring} proposed a static analysis tool, called RefDistiller, that helps developers inspect manual refactoring edits. The tool compares two program versions to detect refactoring anomalies' type and location. It supports six refactoring operations, detects incomplete refactorings, and provides inspection for manual refactorings. 


Coelho \etal  \cite{coelho2019refactoring} performed a systematic literature mapping study on refactoring tools to support modern code review. They raised the need for more tools to explain composite refactorings.  They also reported the need for more surveys to assess the existing refactoring tools for modern code review in both open-source and industrial projects. Pascarella \etal  \cite{pascarella2018information} investigated the effect of code review on bad programming practices (\ie code smells). Their approach mainly focused on comparing code smells at the file level before and after the code review process. Additionally, they manually investigated whether the severity of code smells was reduced in a code review or not. Their results show that in 95\% of the cases, the severity of code smells does not decrease with a review. The reduction in code smells in the remaining few cases was impacted by code insertion and refactoring-related changes. 

Paix{\~a}o \etal  \cite{paixao2020behind} explored if developers’ intents influence the evolution of refactorings during the review of a code change by mining 1,780 reviewed code changes from 6 open-source systems. Their main findings show that refactorings are most often used in code reviews that implement new features, accounting for 63\% of the code changes we studied. Only in 31\% of the code reviews that employed refactorings the developers had the explicit intent of refactoring. Uch{\^o}a \etal \cite{uchoa2020does} reported the multi-project retrospective study that characterizes how the process of design degradation evolves within each review and across multiple reviews. The authors utilized software metrics to observe the influence of certain code review practices on combating design degradation. The authors found that the majority of code reviews had little to no design degradation impact in the analyzed projects. Additionally, the practices of long discussions and the high proportion of review disagreement in code reviews were found to increase design degradation. In their study on predicting design impactful changes in modern code review with technical and/or social aspects, Uch{\^o}a \etal \cite{uchoa2021predicting} analyzed reviewed code changes from seven open source projects. By evaluating six machine learning algorithms, the authors found that technical features result in more precise predictions, and the use of social features alone also leads to accurate predictions.

A couple of studies considered pull requests as a main source of the study code review process. Pantiuchina \etal \cite{pantiuchina2020developers} presented a mining-based study to investigate why developers are performing refactoring in the history of 150 open source systems. Particularly, they analyzed 551 pull requests implemented refactoring operations and reported a refactoring taxonomy that generalizes the ones existing in the literature. Coelho \etal \cite{coelho2021empirical} performed a quantitative and qualitative study exploring code reviewing-related aspects intending to characterize refactoring-inducing pull requests. Their main finding show that refactoring-inducing pull requests take significantly more
time to merge than non-refactoring-inducing pull requests.

AlOmar \etal \cite{alomar2021icse} conducted a case study in an industrial setting to explore refactoring practices in the context of modern code review from the following five dimensions: (1) developers motivations to refactor their code, (2) how developers document their refactoring for code review, (3) the challenges faced by reviewers when reviewing refactoring changes, (4) the mechanisms used by reviewers to ensure the correctness after refactoring, and (5) developers and reviewers assessment of refactoring impact on the source code's quality. Their findings show that refactoring code reviews take longer to be completed than non-refactoring code reviews. In a follow-up work, AlOmar \etal \cite{alomar2022code} performed an emperical study on OpenStack to understand the challenges developers faced when reviewing refactoring changes. Their findings corroborate the results of their industrial case study, indicating that refactoring changes require more time for acceptance compared to non-refactoring changes. 
 Brito and Valente \cite{brito2020refactoring} introduced RAID, a refactoring-aware and intelligent diff tool to alleviate the cognitive effort associated with
code reviews. The tool relied on RefDiff \cite{silva2020refdiff} and is fully integrated with the state-of-the-art practice of continuous integration pipelines (GitHub Actions) and browsers (Google Chrome). The authors evaluated the tool with eight professional developers and found that RAID indeed reduced the cognitive effort required for detecting and reviewing refactorings. In another study, Kurbatova \etal \cite{kurbatova2021refactorinsight} presented RefactorInsight, a plugin for IntelliJ IDEA that integrates information about refactorings in diffs in the IDE, auto folds refactorings in code diffs in Java and Kotlin, and shows hints with their short descriptions.

To summarize, the study of open source projects that use either the Gerrit tools or GitHub pull
requests has been extensively studied (\eg  \cite{rigby2013convergent,thongtanunam2015should,zhang2018multiple, pantiuchina2020developers}). Since notable open-source organizations such as Eclipse, OpenStack, and Qt adopted Gerrit as their code review management tool, we chose to analyze refactoring practices in modern code review from projects that adopted Gerrit as their code review tool. 
Although there are recent studies that explored the motivation behind refactoring in pull requests \cite{pantiuchina2020developers,coelho2021empirical}, to the best of our knowledge, no prior studies have manually extracted all the criteria developers are facing when submitting their refactored code for review from a dedicated `Refactor' branch. To gain a more in-depth understanding of the factors mostly associated with refactoring review and to advance the understanding of refactoring-aware code review, in this paper, we performed an empirical study on a rapidly evolving open-source project. This study complements the existing efforts that are done in an industrial environment \cite{alomar2021icse} and open source systems \cite{pantiuchina2020developers,coelho2021empirical,alomar2022code} using Gerrit and GitHub pull-based development.

\section{Study Design}
\label{Section:Methodology}

The main goal of our study is to understand the practice of refactoring in the context of \textcolor{black}{Modern Code Review (MCR)} to characterize the criteria that influence decision making when reviewing refactoring changes. Thus, we aim to answer the following research questions:

\begin{itemize}
\item \textbf{RQ$_1$.} \textit{\RQone}

\item \textbf{RQ$_2$.} \textit{\RQtwo}

\item \textbf{RQ$_3$.} \textit{\RQthree}

\item \textbf{RQ$_4$.} \textit{\RQfour}
\end{itemize}

\begin{sidewaysfigure}[htbp]
 	\centering
 	\includegraphics[width=1.0\textwidth]{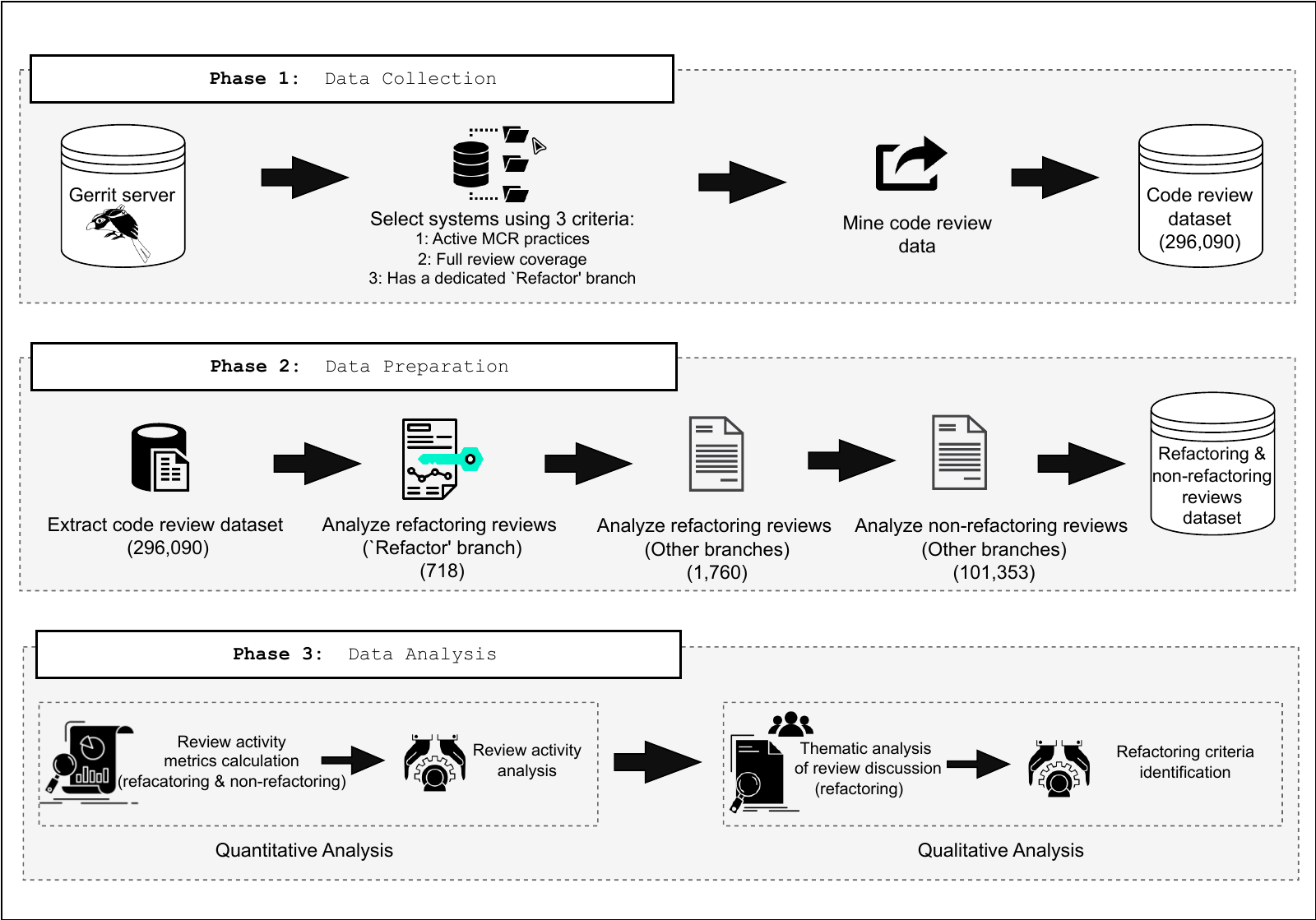}
 	\caption{Overview of our experiment design.}
 	\label{fig:approach}
\end{sidewaysfigure}

According to the guidelines reported by Runeson and H{\"o}st \cite{runeson2009guidelines}, we designed an empirical study that consists of three steps, as depicted in Figure \ref{fig:approach}, and discussed in the next subsections. 
  Since our research questions are both quantitative and qualitative, we used tools/scripts along with manual activities to investigate our data. 
Furthermore, the dataset utilized in this study is available on our project website \cite{ReplicationPackage} for extension and replication purposes.


\subsection{Data Collection}
\subsubsection{Studied Systems}
\textcolor{black}{In line with \cite{thongtanunam2015investigating,thongtanunam2016revisiting,morales2015code}}, to select the subject systems, we identified three important criteria:

\noindent \textbf{Criterion \#1: Active MCR practices.} Our goal is to study a system that actively examines code changes through a code review tool. Therefore, we focus on systems where a number of reviews are performed using a code review tool \textcolor{black}{(\ie systems which have review procedures in place)}, similar to \cite{thongtanunam2015investigating,thongtanunam2016revisiting,morales2015code}. 

\noindent \textbf{Criterion \#2: Full review coverage.}
Since we investigate the practice of refactoring-related code reviews, we focus on systems that have many files with 100\% review coverage \textcolor{black}{(\ie files where every change made to them is reviewed before they are merged into the repositories)}, similar to studies that explored code review practices in defective files \cite{thongtanunam2015investigating,thongtanunam2016revisiting,mcintosh2014impact}. 

\noindent \textbf{Criterion \#3: Has a dedicated `Refactor' branch.} Since we want to study refactoring practices in MCR, we need to ensure that the subject systems have sufficient refactoring-related instances to help us perform our statistical analysis. \textcolor{black}{So, we selected the project with the highest number of refactoring reviews.}


\textcolor{black}{To satisfy criterion 1, we started by considering five systems (\ie  OpenStack,\footnote{https://review.opendev.org/} Qt,\footnote{ https://codereview.qt-project.org/} LibreOffice,\footnote{https://gerrit.libreoffice.org/} VTK,\footnote{http://vtk.org/} ITK\footnote{http://itk.org/}) 
that use the Gerrit code review tool and have been widely studied in previous research
in MCR, \eg \cite{thongtanunam2020review,ouni2016search,fan2018early,chouchen2021whoreview}. We then discarded VTK and ITK since Thongtanunam \etal \cite{thongtanunam2016revisiting} reported that the linkage rate of code changes to the reviews for VTK is too low and ITK does not satisfy criterion 2.  As for criterion 3, after mining the code review data, we found that Qt is the only system with a `Refactor' branch. Due to the human-intensive nature of carefully studying and analyzing refactoring practice in MCR, we opt for performing an in-depth study on a single system. With the above-mentioned criteria in mind, we select Qt, 
 a cross-platform application and user interface framework developed by Digia Corporation.}

\subsubsection{Mining Code Review Data}
We mined code review data using the \texttt{RESTful API}\footnote{https://gerrit-review.googlesource.com/Documentation/rest-apichanges.html} provided by Gerrit, which returns the results in a \texttt{JSON} format. We used a script to automatically mine the review data and store them in the \texttt{SQLite} database. All collected reviews are closed (\ie  having a status of either \say{\textit{Merged}} or \say{\textit{Abandoned}}). 
In total, we mined 296,372 code changes between December 2012 and April 2021 from Qt projects. An overview of the project's statistics is provided in Table~\ref{Table:DATA_Overview}.
\begin{table}[h!]
\begin{center}
\caption{Overview of the Qt studied system.} 
\label{Table:DATA_Overview}
\begin{adjustbox}{width=0.7\textwidth,center}
\rowcolors{2}{white}{gray!25}
\begin{tabular}{lllll}\hline
\toprule
\bfseries Item & \bfseries Count \\
\midrule
Version & 4.7 to 5.11 \\
Line of code & 21,256,665 \\
No. of commits & 1,659,190 \\
No. of code changes &  296,372 \\
No. of developers & 3,264  \\
No. of files &  351,387  \\
Reviews in `Refactor' branch & 718 \\
Reviews with keyword `\textit{refactor*}' in title and description & 1,760  \\
\textcolor{black}{Non-refactoring reviews from other branches} & \textcolor{black}{101,353}\\
\bottomrule
\end{tabular}
\end{adjustbox}
\end{center}
\end{table}

\subsection{Data Preparation}

Our main goal is to explore refactoring review \textit{culture} in `Refactor' branch. However, to make a comparison, we select refactoring reviews containing the keyword `\textit{refactor}'. Similarly to previous work on identifying refactoring changes or defect-fixing or defect-inducing changes \cite{kim2008classifying,kamei2012large,mockus2000identifying,hassan2008automated,thongtanunam2016revisiting,mcintosh2016empirical,pantiuchina2020developers,Ratzinger:2008:RRS:1370750.1370759,coelho2021empirical,alomar2019can,alomar2021ESWA,stroggylos2007refactoring, alomar2019impact,tang2021empirical}, we utilize a keyword-based mechanism to extract refactoring code review data from other branches. The keyword-based approach was chosen for the manual inspection, which required not only non-trivial efforts but also a deep knowledge of the domain. Specifically, we start by searching for the term `\textit{refactor*}' in the title or description (we use * to capture extensions like refactors, refactoring etc.). The choice of `\textit{refactor}', besides being used by various related studies, is intuitively the first term to identify refactoring-related code review. However, since related work on refactoring documentation shows that developers may use other synonymous terms/phrases \cite{alomar2021documentation,alomar2021ESWA,Ratzinger:2008:RRS:1370750.1370759,zhangpreliminary18}, we ensure to exclude these synonymous terms/phrases when selecting non-refactoring reviews. In summary, we have extracted the following reviews.

\begin{itemize}
  \item \textit{Refactoring reviews from the `Refactor' branch.} These reviews are specifically chosen from a dedicated `Refactor' branch.
\item \textit{Refactoring reviews containing the keyword `\textit{refactor}'.} These reviews include the keyword `\textit{refactor}' in their title and description and are selected from branches other than the `Refactor' branch.
\item \textit{Non-refactoring reviews from other branches.} These reviews are selected from branches other than the `Refactor' branch, and they do not involve refactoring or any synonymous terms/phrases commonly found in the literature.
\end{itemize}

To extract the set of refactoring-related code reviews, we follow a two-step procedure: (1) automatic filtering, and (2) manual filtering. 

\textbf{(1) Automatic Filtering.} \textcolor{black}{In the first step, we extract all of the 718 review instances in the `Refactor' branch.  We notice that the ratio of these reviews is very small in comparison with the total number of the mined reviews, \ie  296,372.}

\textbf{(2) Manual Filtering.} To ensure the correctness of the data, we manually inspected and read all these refactoring reviews. 
 Our goal is to have a \textit{gold set} of reviews in which the developers explicitly reported the refactoring activity. This \textit{gold set} will serve to check later criteria that are mostly associated with refactoring review discussion. \textcolor{black}{Furthermore, since related work on refactoring documentation shows that developers may use synonymous terms/phrases \cite{alomar2021documentation,alomar2021ESWA,Ratzinger:2008:RRS:1370750.1370759,zhangpreliminary18}, we ensure to exclude these synonymous terms/phrases and manually inspect them when selecting non-refactoring reviews.}
 




\subsection{Data Analysis}
To address our research questions, a structured mixed-method study was designed to combine elements of both quantitative and qualitative research.

\subsubsection{Quantitative data analysis.} We leverage the data collected to compare refactoring and non-refactoring reviews using review efforts, \ie code review metrics.
 As we calculate the metrics of refactoring and non-refactoring code reviews, we want to distinguish, for each metric, whether the variation is statistically significant. 
 We first test for normality using the Shapiro-Wilk normality test \cite{taeger2014statistical} and observe that the distribution of code review activity metrics does not follow a normal distribution. Therefore, we use the Mann-Whitney U test \cite{conover1998practical}, a non-parametric test, to compare between the two groups, since these groups are independent of one another. The null hypothesis is defined by no variation in the metric values of refactoring and non-refactoring code reviews. Thus, the alternative hypothesis indicates that there is a variation in the metric values. Additionally, the variation between values of both sets is considered significant if its associated \textit{p}-value is less than 0.05. Furthermore, we use the Cliff's Delta ($\delta$) \cite{cliff1993dominance}, a non-parametric effect size measure, to estimate the magnitude of the differences between refactoring and non-refactoring reviews. As for its interpretation, we follow the guidelines reported by Romano \etal \cite{romano2006appropriate}:
 
 \begin{itemize}
\item Negligible for $\mid \delta \mid< 0.147$
\item Small for $0.147 \leq \mid \delta \mid < 0.33$
\item Medium for $0.33 \leq \mid \delta \mid < 0.474$
\item Large for $\mid \delta \mid \geq 0.474$
\end{itemize}
 
 To measure the extent of the relationship between these metrics, we conducted a Spearman rank correlation test (a non-parametric measure) \cite{wissler1905spearman}. We chose a rank correlation because this type of correlation is resilient to data that is not normally distributed. 

\subsubsection{Qualitative data analysis.} To answer RQ$_2$, RQ$_3$, and RQ$_4$, we perform the analysis of the data. The author manually inspects refactoring review subject, description, and discussions \textcolor{black}{by considering both the general comments and the inline comments}. 
  Next, we describe the methodology for building and refining the taxonomy. 

\vspace{.1cm}
{\textbf{Taxonomy Building and Refinement.}} \textcolor{black}{The goal of the manual analysis was to categorize the topics discussed in the ‘Refactor’ branch within each of the refactoring review instances. The entire process was supported by a spreadsheet application
equipped with tagging capabilities. For each instance, the
evaluator was presented with: (i) the metadata as returned by
Gerrit (\eg Gerrit\_Id, Gerrit\_URL); (ii) the branch that was matched in that specific instance; and (iii) the subject and description in Gerrit for easier inspection.}

\textcolor{black}{The categorization required the assignment of one or more labels to an instance, describing the topics discussed. In case manual inspection revealed
that reviews were not actually used for refactoring
tasks, the instance was discarded.}

When analyzing the review discussions, we adopted a thematic analysis approach based on the guidelines provided by Cruzes \etal \cite{cruzes2011recommended}. Thematic analysis is one of the most used methods in Software Engineering literature (\eg \cite{Silva:2016:WWR:2950290.2950305}), which is a technique for identifying and recording patterns (or \say{themes}) within a collection of descriptive labels, which we call \say{codes}. For each refactoring review, we proceeded with the analysis using the following steps: 
i) Initial reading of the review discussions; ii) Generating initial codes (\ie labels) for each review; iii) Translating codes into themes, sub-themes, and higher-order themes; iv) Reviewing the themes to find opportunities for merging; v) Defining and naming the final themes, and creating a model of higher-order themes and their underlying evidence.

\textcolor{black}{The above-mentioned steps were performed independently by two annotators. One annotator performed the labeling of review discussions independently of the other author who was responsible for reviewing the taxonomy currently drafted. By the end of each iteration, the authors met and refined the taxonomy}. 

It is important to note that the approach is not a single-step process. As the codes 
 were analyzed, some of the first cycle codes 
were subsumed by other codes, relabeled, or dropped altogether. As the author progressed with the translation to themes, there was some rearrangement, refinement, and reclassification of data into different or new codes. For example, we aggregated, into \say{\textit{Refactoring}}, the preliminary categories \say{\textit{move method}}, \say{\textit{refactoring operations}}, and \say{\textit{rename}} that were analyzed. We used the thematic analysis technique to address RQ$_4$.


\textcolor{black}{\textbf{\textcolor{black}{Taxonomy Validation.}} In addition to the iterative process of building the taxonomy, we need to externally validate it from a practitioner's point of view \cite{pascarella2018self,dougan2022towards}. The aim of this validation is to investigate whether it reflects actual MCR practices. To do so, we validated the taxonomy with a senior developer, with 8 years of industrial experience, and with 4 years of experience in code review. The survey contained 9 questions related to the correctness and representativeness of our taxonomy and proposed guidelines.} 

\section{Results and Discussion}
\label{Section:Result}

\subsection{\RQone} 



\begin{sidewaysfigure}
\centering
\begin{subfigure}{6cm}
\centering\includegraphics[width=6cm]{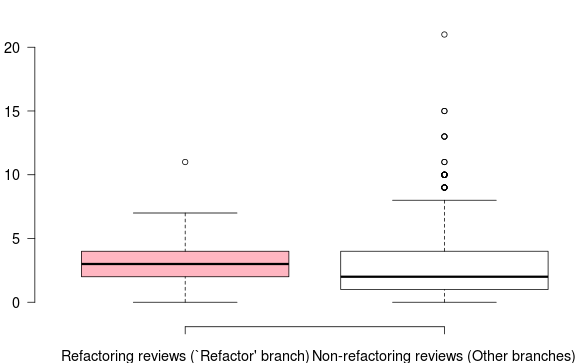}
\caption{Number of reviewers}
\label{BP:reviewers2}
\end{subfigure}%
\begin{subfigure}{6cm}
\centering\includegraphics[width=6cm]{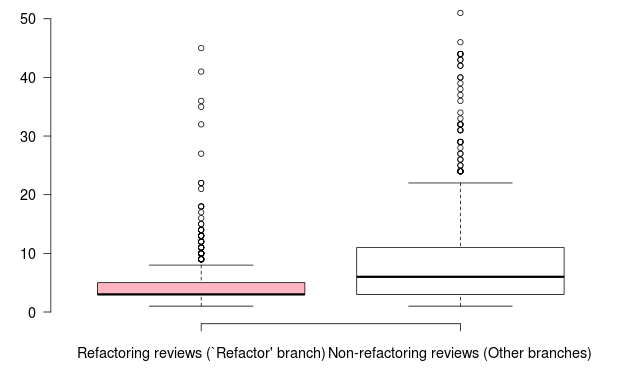}
\caption{Number of review comments}
\label{BP:messages2}
\end{subfigure}%
\begin{subfigure}{6cm}
\centering\includegraphics[width=6cm]{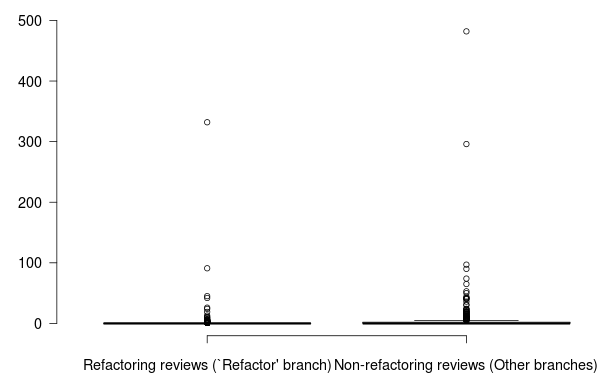}
\caption{Number of inline comments}
\label{BP:inlinecomments2}
\end{subfigure}

\begin{subfigure}{6cm}
\centering\includegraphics[width=6cm]{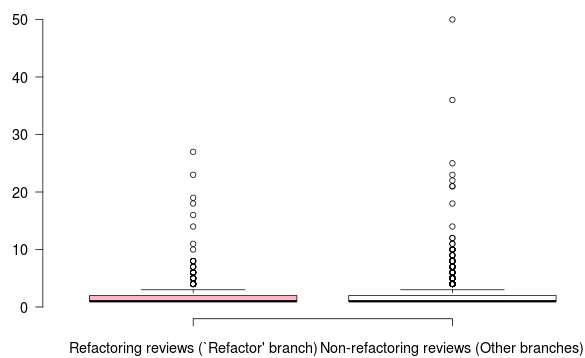}
\caption{Number of revisions}
\label{BP:revisions2}
\end{subfigure}%
\begin{subfigure}{6cm}
\centering\includegraphics[width=6cm]{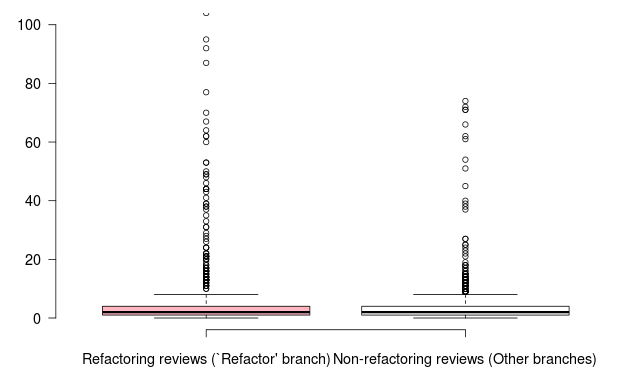}
\caption{Number of changed files}
\label{BP:files2}
\end{subfigure}%
\begin{subfigure}{6cm}
\centering\includegraphics[width=6cm]{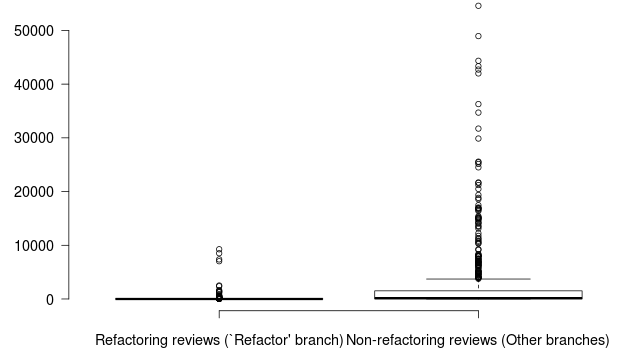}
\caption{Review duration (seconds)}
\label{BP:duration2}
\end{subfigure}%

\begin{subfigure}{6cm}
\centering\includegraphics[width=6cm]{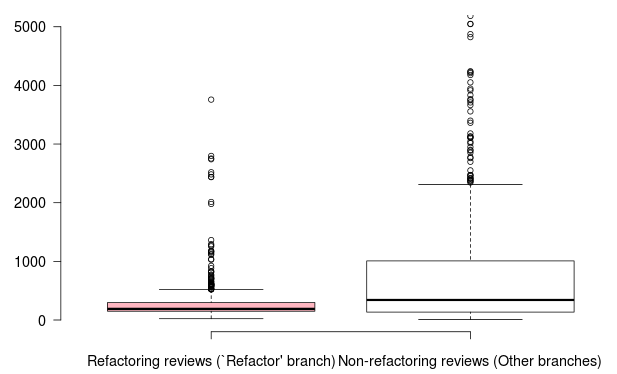}
\caption{Length of discussion (characters)}
\label{BP:discussion2}
\end{subfigure}%
\begin{subfigure}{6cm}
\centering\includegraphics[width=6cm]{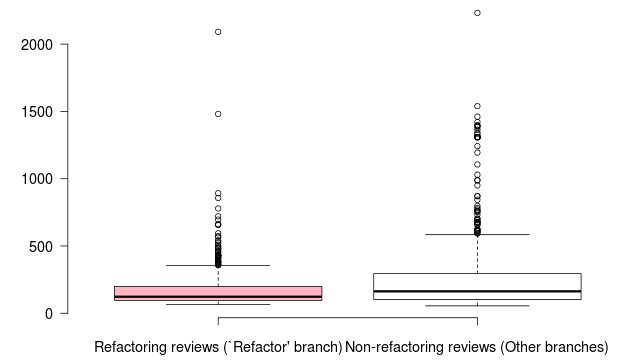}
\caption{Length of description (characters)}
\label{BP:description2}
\end{subfigure}%
\begin{subfigure}{6cm}
\centering\includegraphics[width=6cm]{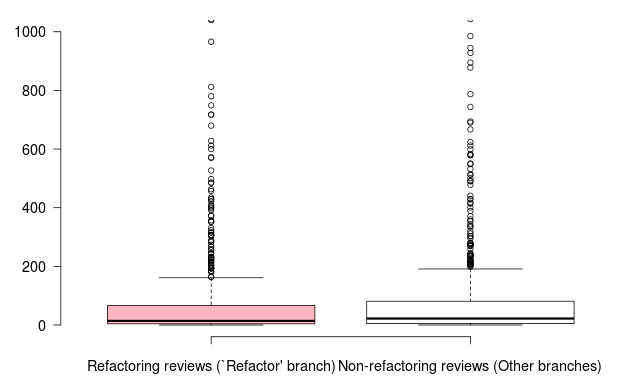}
\caption{Code churn}
\label{BP:Churn2}
\end{subfigure}%

\caption{Boxplots of metrics values of refactoring reviews (`Refactor') branch and non-refactoring reviews (Other branches).} 
\label{Chart:Boxplots_2}
\end{sidewaysfigure}

\begin{sidewaysfigure}
\centering
\begin{subfigure}{6cm}
\centering\includegraphics[width=6cm]{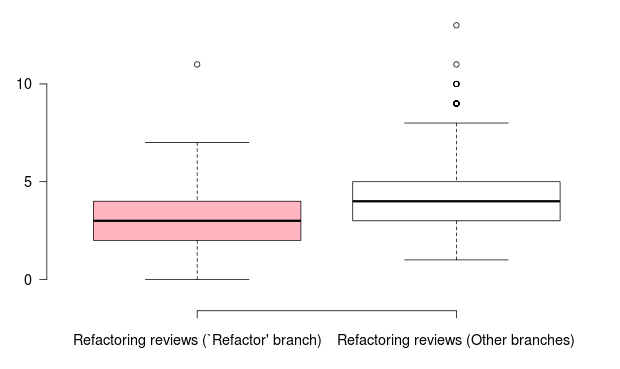}
\caption{Number of reviewers}
\label{BP:reviewers1}
\end{subfigure}%
\begin{subfigure}{6cm}
\centering\includegraphics[width=6cm]{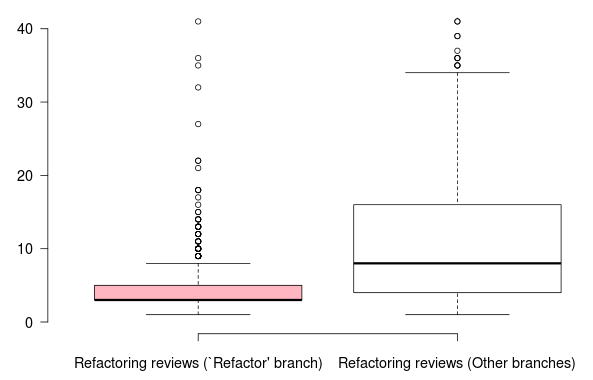}
\caption{Number of review comments}
\label{BP:messages1}
\end{subfigure}%
\begin{subfigure}{6cm}
\centering\includegraphics[width=6cm]{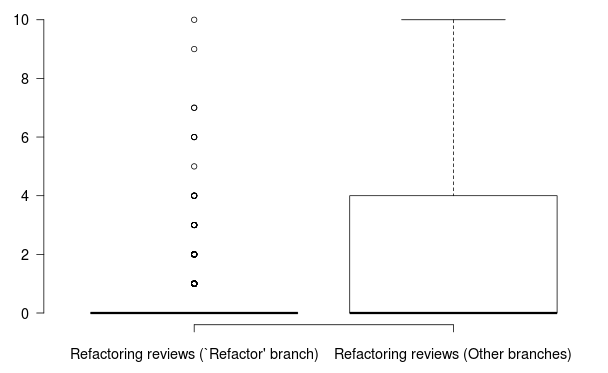}
\caption{Number of inline comments}
\label{BP:inlinecomments1}
\end{subfigure}

\begin{subfigure}{6cm}
\centering\includegraphics[width=6cm]{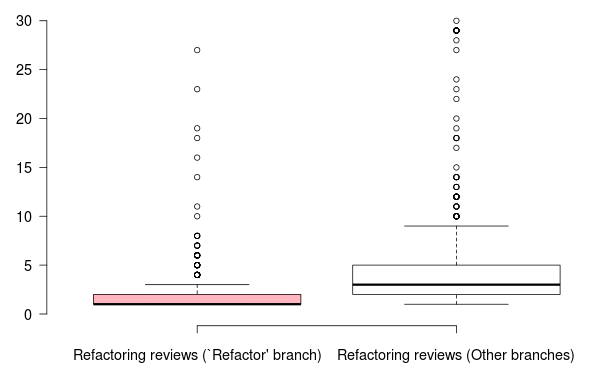}
\caption{Number of revisions}
\label{BP:revisions1}
\end{subfigure}%
\begin{subfigure}{6cm}
\centering\includegraphics[width=6cm]{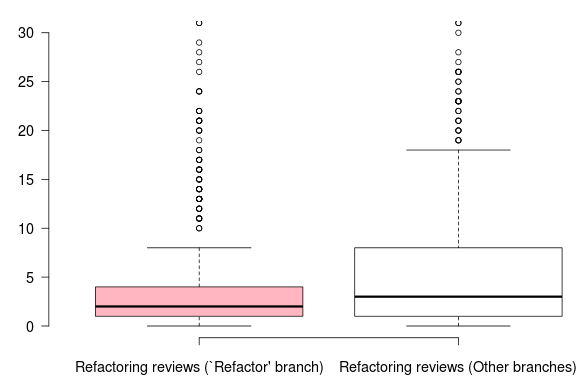}
\caption{Number of changed files}
\label{BP:files1}
\end{subfigure}%
\begin{subfigure}{6cm}
\centering\includegraphics[width=6cm]{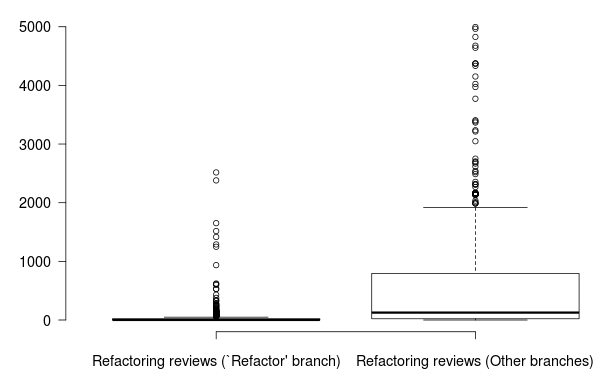}
\caption{Review duration (seconds)}
\label{BP:duration1}
\end{subfigure}%

\begin{subfigure}{6cm}
\centering\includegraphics[width=6cm]{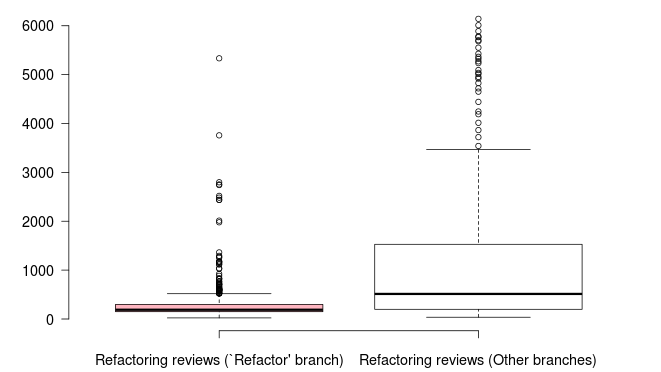}
\caption{Length of discussion (characters)}
\label{BP:discussion1}
\end{subfigure}%
\begin{subfigure}{6cm}
\centering\includegraphics[width=6cm]{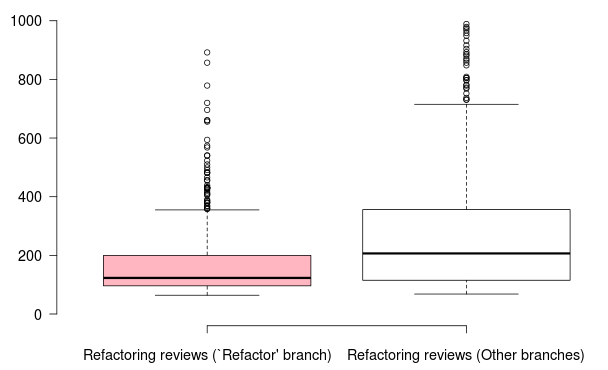}
\caption{Length of description (characters)}
\label{BP:description1}
\end{subfigure}%
\begin{subfigure}{6cm}
\centering\includegraphics[width=6cm]{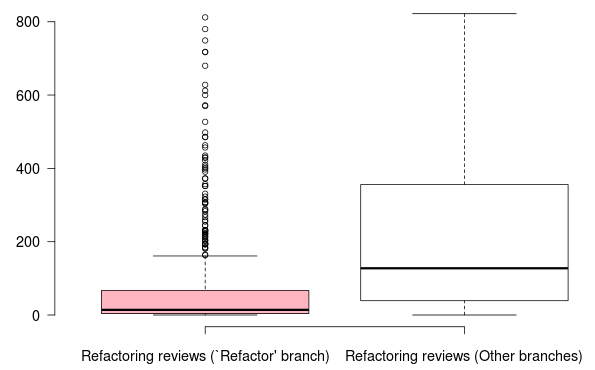}
\caption{Code churn}
\label{BP:Churn1}
\end{subfigure}%

\caption{Boxplots of metrics values of refactoring reviews (`Refactor') branch and refactoring reviews (Other branches).} 
\label{Chart:Boxplots_1}
\end{sidewaysfigure}

\noindent\textbf{Motivation.} \textcolor{black}{The first research question aims to explore whether reviewing refactoring in `Refactor' branch takes longer compared to refactoring reviews containing the keyword `refactor' and non-refactoring reviews from other branches. Understanding the differences in review efforts helps identify the unique challenges and requirements of refactoring reviews compared to non-refactoring ones.}
\noindent\textbf{Approach.} \textcolor{black}{To address RQ1, we intend to compare \textit{refactoring reviews} with \textit{non-refactoring reviews}, to see whether there are any differences in terms of code review efforts or metrics listed in Table \ref{Table:RQ1_results}, and Figures \ref{Chart:Boxplots_2}, \ref{Chart:Boxplots_1}. 
Since our refactoring set in the `Refactor' branch contains 718 reviews, we sampled 718 non-refactoring reviews and refactoring reviews containing the keyword `refactor*' from the remaining ones in the review framework. This size provides a comprehensive view of the refactoring review practices within the `Refactor' branch, capturing a diverse range of scenarios and developer interactions. To ensure the representativeness of the sample \cite{clarkson1989applications}, we use stratified random sampling by choosing reviews from the rest of the reviews.}

\begin{table*}
\centering
\caption{\textcolor{black}{Statistics of code review activity efforts.}} 
\label{Table:RQ1_results}
\begin{sideways}
\begin{adjustbox}{width=1.1\textwidth,center}
\begin{tabular}{@{}lllllll|llllll|ll@{}}
\toprule
\multicolumn{1}{c}{\multirow{3}{*}{\textbf{Metrics}}} & \multicolumn{6}{c|}{\textit{\textbf{Refactoring code review (`refactor' branch)}}} & \multicolumn{6}{c|}{\textit{\textbf{Non-refactoring code review}}} & \multicolumn{2}{c}{\textit{\textbf{Statistical difference}}}  \\ \cmidrule(l){2-15} 
\multicolumn{1}{c}{} & \multicolumn{1}{c}{\textbf{Min}} & \multicolumn{1}{c}{\textbf{Q1}} & \multicolumn{1}{c}{\textbf{Median}} & \multicolumn{1}{c}{\textbf{Mean}} & \multicolumn{1}{c}{\textbf{Q3}} & \multicolumn{1}{c|}{\textbf{Max}} & \multicolumn{1}{c}{\textbf{Min}} & \multicolumn{1}{c}{\textbf{Q1}} & \multicolumn{1}{c}{\textbf{Median}} & \multicolumn{1}{c}{\textbf{Mean}} & \multicolumn{1}{c}{\textbf{Q3}} & \multicolumn{1}{l|}{\textbf{Max}} &
\multicolumn{1}{c}{\textbf{\textit{p}-value}} & \multicolumn{1}{l}{\textbf{Cliff's delta ($\delta$)}}\\ 
\midrule
Number of reviewers & 0 & 2 & 3 &  2.93 & 4  & 7 & 0 &1 &2  &3.01  &4  & 8 &  \textbf{0} & small (0.11)\\
Number of review comments & 1 & 3 & 3 & 4.67 & 5 & 8 & 1 & 3 &6 & 9.20  & 11 & 22 & \textbf{0} & medium (0.3) \\
Number of inline comments & 0 & 0 & 0 & 1.12 & 0 & 0 & 0 & 0 & 0 & 3.72 & 2 & 5 & \textbf{0} & medium (0.34) \\
Number of revisions & 1 & 1 & 1 & 1.87 & 2 & 3 & 1 & 1 & 1 & 2.44 & 2 & 3 & \textbf{0.000211} & small (0.09)\\
Number of changed files & 0 & 1 & 2 & 49.84 & 4 & 8 & 0 & 1 & 2 & 9.08 & 4 & 8 &  0.6369 & small (0.01)  \\
Review duration (seconds) & 0 & 0.15  & 1.08 & 90.58 & 20.66 & 48.41 & 0 & 16.96 & 163.67 & 2356.32 & 1515.01 & 3719.12 & \textbf{0} & medium (0.5) \\
Length of discussion (characters)   & 22 & 150 & 190 & 427.32 & 298 & 518 & 9 & 135 & 341.50 & 2160.06 & 1008 & 2312 &  \textbf{1.621e-14} & small (0.2)\\
Length of description (characters) & 64 & 96 & 123 & 172.12 & 200 & 355 & 55 & 102 & 163 & 260.2 & 295 & 584 & \textbf{1.239e-8} & small (0.15) \\
Code churn  & 0 & 4 & 14 & 1366.49 & 67 & 161 & 0 & 5 & 22 & 364.82 & 81 & 191 & \textbf{0.005045} &  small (0.07) \\ \hline
\multicolumn{1}{c}{\multirow{3}{*}{\textbf{Metrics}}} & \multicolumn{6}{c|}{\textit{\textbf{Refactoring code review (`refactor' branch)}}} & \multicolumn{6}{c|}{\textit{\textbf{Refactoring code review (Other branches)}}} & \multicolumn{2}{c}{\textit{\textbf{Statistical difference}}}  \\ \cmidrule(l){2-15} 
\multicolumn{1}{c}{} & \multicolumn{1}{c}{\textbf{Min}} & \multicolumn{1}{c}{\textbf{Q1}} & \multicolumn{1}{c}{\textbf{Median}} & \multicolumn{1}{c}{\textbf{Mean}} & \multicolumn{1}{c}{\textbf{Q3}} & \multicolumn{1}{c|}{\textbf{Max}} & \multicolumn{1}{c}{\textbf{Min}} & \multicolumn{1}{c}{\textbf{Q1}} & \multicolumn{1}{c}{\textbf{Median}} & \multicolumn{1}{c}{\textbf{Mean}} & \multicolumn{1}{c}{\textbf{Q3}} & \multicolumn{1}{l|}{\textbf{Max}} &
\multicolumn{1}{c}{\textbf{\textit{p}-value}} & \multicolumn{1}{l}{\textbf{Cliff's delta ($\delta$)}}\\ 
\midrule
Number of reviewers & 0 & 2 & 3 &  2.93 & 4  & 7 & 1 &3 &4  &3.82  &5 & 8 &  \textbf{0} & small (0.27)\\
Number of review comments & 1 & 3 & 3 & 4.67 & 5 & 8 & 1 & 4 &8 & 12.84  & 16 & 34 & \textbf{0} & medium (0.43) \\
Number of inline comments & 0 & 0 & 0 & 1.12 & 0 & 0 & 0 & 0 & 0 & 5.55 & 4 & 10 & \textbf{0} & medium (0.38) \\
Number of revisions & 1 & 1 & 1 & 1.87 & 2 & 3 & 1 & 2 & 3 & 4.13 & 5 & 9 & \textbf{0} & medium (0.45)\\
Number of changed files & 0 & 1 & 2 & 49.84 & 4 & 8 & 0 & 1 & 3 & 7.82 & 8 & 18 &  \textbf{2.187e-14} & small (0.2)  \\
Review duration (seconds) & 0 & 0.15  & 1.08 & 90.58 & 20.66 & 48.41 & 0 & 21.37 & 126.09 & 1342.24 & 1917.12 & 3719.12 & \textbf{0} & large (0.56) \\
Length of discussion (characters)   & 22 & 150 & 190 & 427.32 & 298 & 518 & 36 & 198 & 513 & 6571.11 & 1528 & 3469 &  \textbf{0} & medium (0.37)\\
Length of description (characters) & 64 & 96 & 123 & 172.12 & 200 & 355 & 68 & 115 & 206.5 & 304.15 & 356 & 715 & \textbf{0} & medium (0.31) \\
Code churn  & 0 & 4 & 14 & 1366.49 & 67 & 161 & 0 & 39 & 127.5 & 425.26 & 356 & 822 & \textbf{0} &  medium (0.44) \\ 
\bottomrule
\end{tabular}
\end{adjustbox}
\end{sideways}
\end{table*}
\newpage
\noindent\textbf{Results.} By looking at the statistical summary in Table \ref{Table:RQ1_results}, and Figures \ref{Chart:Boxplots_2}, \ref{Chart:Boxplots_1}, we found that reviewing refactoring changes in `Refactor' branch differs, with fewer reviewers ($\mu$ = 2.93), fewer review comments ($\mu$ = 4.67), fewer inline comments ($\mu$ = 1.12), fewer revisions ($\mu$ = 1.87), shorter review time ($\mu$ = 90.58), and fewer discussions and descriptions ($\mu$ = 427.32, $\mu$ = 172.12, respectively) compared to reviewing non-refactoring changes. However,  reviewing refactoring changes in the `Refactor' branch shows more file changes ($\mu$ = 49.84), and more added and deleted lines between revisions ($\mu$ = 1366.49). As shown in Table \ref{Table:RQ1_results}, we performed a non-parametric Mann-Whitney U test and we obtained a statistically significant \textit{p}-value when the values of these two groups were compared (\textit{p}-value $< 0.05$ for all review efforts, except number of changed files), and accompanied with a small, medium, or large effect size depending on the review effort/metric. 

Regarding the comparison between reviewing refactoring changes from the `Refactor' branch and reviewing refactoring changes with the keyword `refactor*' from other branches, we observed similar patterns as in the previous comparison. Reviewing refactoring changes in the `Refactor' branch significantly differs, with fewer reviewers, fewer review comments, fewer inline comments, fewer revisions, shorter review time, and fewer discussions and descriptions compared to reviewing refactoring changes from other branches. However, in this comparison, there are more changed files and more added and deleted lines between revisions when reviewing refactoring changes from other branches. The only difference between this set and the previous comparison is that the difference in the number of changed files is significant.




\textcolor{black}{We speculate that the observed differences in reviewing refactoring changes from the `Refactor' branch and reviewing refactoring changes with the keyword `refactor*' from other branches and non-refactoring reviews can be attributed to several factors:} 

\begin{figure*}[htbp]
\centering 
\includegraphics[width=\textwidth]{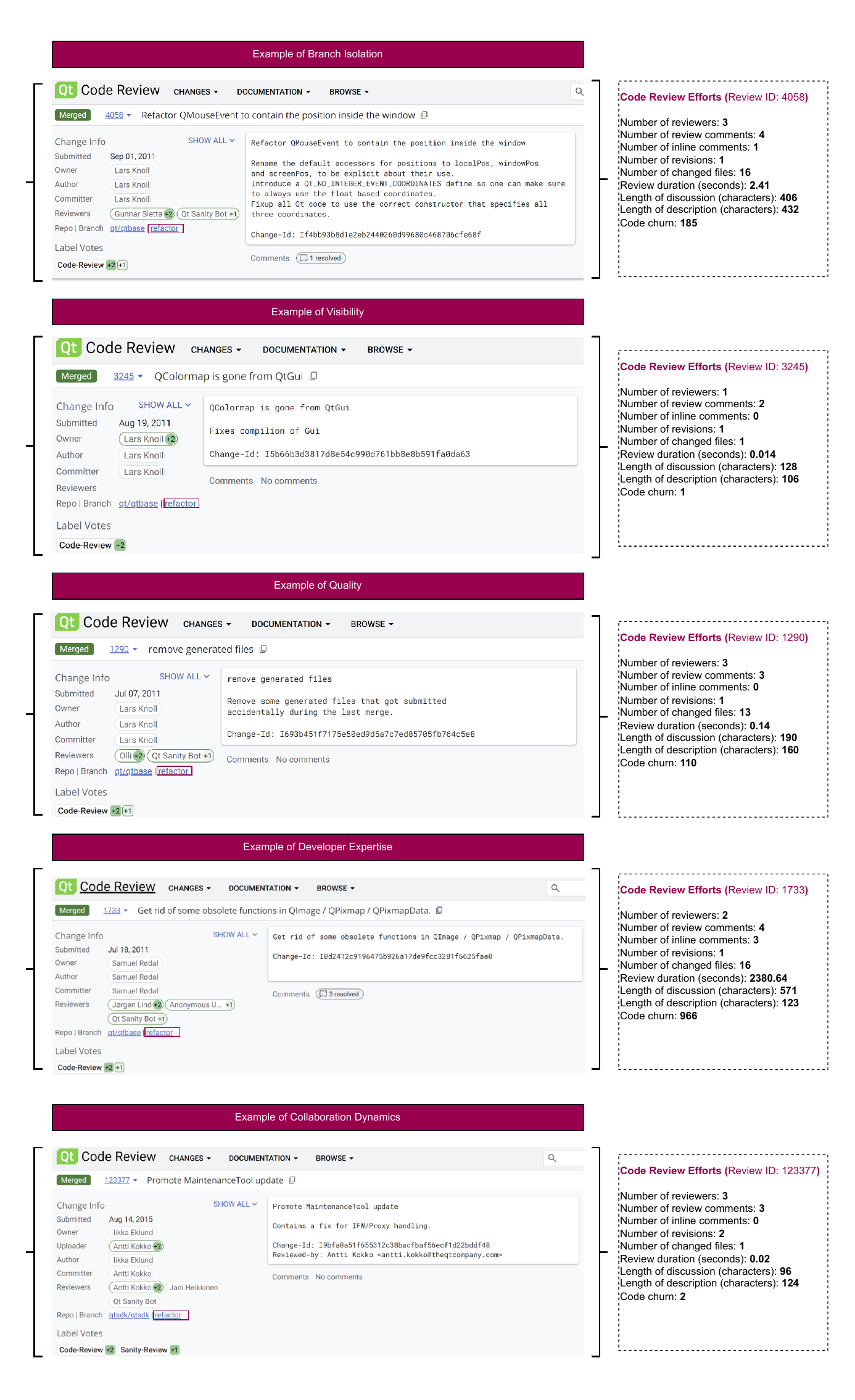}
\caption{\textcolor{black}{Example of refactoring reviews from the `Refactor' branch in the Qt project.}}
\label{fig:example-casestudies}
\end{figure*}

\noindent\textcolor{black}{\textbf{Branch Isolation.} Reviewing refactoring changes in the `Refactor’ branch took significantly less time compared to the non-refactoring changes. This indicates that the isolation of refactoring activities could streamline the review process, as reviewers can focus solely on refactoring tasks without being distracted by other types of changes. The statistically significant shorter review time (\textit{p}-value $<$ 0.05) in the `Refactor' branch compared to other branches supports the idea that a dedicated branch for refactoring helps reduce the overall review effort.} 
\begin{itemize}
    \item \textcolor{black}{Example. In review ID 4058 (see Figure \ref{fig:example-casestudies}), the refactoring change was reviewed in just 2.41 seconds with only 4 comments, highlighting how the isolated branch facilitated a quick and focused review.}
\end{itemize}
\textcolor{black}{\noindent\textbf{Visibility.} Fewer review comments and inline comments in the `Refactor' branch suggest that reviewers quickly understand the changes and provide focused feedback. This lower number of comments, supported by a statistically significant \textit{p}-value, implies that changes in the `Refactor' branch might be receiving more targeted and effective attention from developers specializing in refactoring.} 
\begin{itemize}
    \item \textcolor{black}{Example. Review ID 3245 (see Figure \ref{fig:example-casestudies}) had 0 inline comments and was approved by a single reviewer, showing how dedicated visibility in the `Refactor' branch helps in quick and clear review processes.}
\end{itemize}
 \textcolor{black}{\noindent\textbf{Quality.} The number of revisions needed and fewer discussions and descriptions in the `Refactor' branch compared to other branches indicate a higher initial code quality. The lower number of revisions and the fewer extensive discussions, with statistically significant differences, suggest that the refactoring changes are better prepared and understood, aligning with the higher quality hypothesis.} 
\begin{itemize}
    \item \textcolor{black}{Example. Review ID 1290 (see Figure \ref{fig:example-casestudies}) required just one revision and had minimal back-and-forth discussion, suggesting that the initial refactoring proposal was of high quality.}
\end{itemize}
\textcolor{black}{\noindent\textbf{Developer Expertise.} A lower number of reviewers is required for refactoring changes in the `Refactor' branch compared to non-refactoring changes. The significant difference in the number of reviewers, supported by \textit{p}-value $<$ 0.05, indicates that specialized knowledge in the `Refactor' branch allows for more efficient and effective reviews, with fewer people needed to reach a consensus.}
\begin{itemize}
    \item \textcolor{black}{Example. In review ID 1733 (see Figure \ref{fig:example-casestudies}), two experienced developers quickly approved the change with minimal comments, demonstrating the impact of specialized expertise on the efficiency of the review process.}
\end{itemize}
\textcolor{black}{\noindent\textbf{Collaboration Dynamics.} Concise discussions and descriptions in the `Refactor' branch suggest a more focused and collaborative environment. The statistically significant reduction in discussion length supports the notion that a collaborative environment in the `Refactor' branch leads to more efficient communication and quicker resolution of review comments.} 
\begin{itemize}
    \item \textcolor{black}{Example. Review ID 123377 (see Figure \ref{fig:example-casestudies}) had a concise discussion with 96 characters and was finalized quickly in 0.22 seconds, reflecting the collaborative and focused nature of the `Refactor' branch.}
\end{itemize}

Moreover, we conjecture that reviewing refactoring from other branches triggers longer discussions between the code change authors and the reviewers as we notice that several refactoring-related actions are being extensively discussed before reaching an agreement. While previous studies have found a similar pattern in GitHub's pull requests in open-source systems \cite{coelho2021empirical}, Gerrit \cite{alomar2022code}, and using code review tools in industry \cite{alomar2021icse}, there is no study that looked at the main reasons for refactoring-related discussions in `Refactor' branch to take significantly less effort to be reviewed. Therefore, the findings of RQ1 have motivated us to manually analyze these reviews and extract the main criteria related to reviewing refactored code (RQ$_4$). 

Further, we observe that refactoring-related code reviews from `Refactor' branch impact larger code churn and more changes across files than non-refactoring code changes. These results are expected and agree with previous work \cite{coelho2021empirical,hegedHus2018empirical,paixao2019impact}, which found that refactored code has higher size-related metrics and larger changes promote refactorings. We also noticed that the number of developers who participated in the code review process is lower due to the high number of lines added, modified, or deleted between revisions. However, unlike a previous finding \cite{coelho2021empirical}, no evidence of the correlation between the number of reviewers and refactoring was detected. 

\begin{boxK}
\textit{\textbf{Summary.} \textcolor{black}{In the `Refactor' branch, the review process is more focused due to the branch's dedicated purpose. This allows developers to concentrate more on refactoring activities. Further, reviews in the `Refactor' branch tend to have shorter review times. This can be attributed to the specialized knowledge of the reviewers who are familiar with refactoring techniques, leading to quicker consensus and fewer iterations.}
}
\end{boxK}


\begin{table}
\centering
\caption{\textcolor{black}{List of refactoring documentation (`*' captures the extension of the keyword).}}
\label{Table:GeneralPatterns}
\small
\begin{adjustbox}{width=0.7\textwidth,center}

\begin{tabular}{lllllll}
\toprule
\textbf{Patterns}\\
\midrule
\textit{`Refactor' branch} \\  \hline \\
Chang* (890) & Fix* (340) & Add* (226) & Mov* (226) & Creat* (139) \\
Remov* (121) & Refactor* (107) & Merg* (74) & Renam* (33) &  Dependenc* (28) \\
Replac* (24) & Get rid of (24) & Improv* (18) & Introduc* (18) & Cleanup* (13) \\
Modif* (6) & Split* (5) & Extend* (4) & Extract* (4) & Polish* (4) \\
Reduc* (4)  & Remov* unused (4) & Inlin* (3) & Simplif* (3) & Encapsulat* (2)\\
 Code clean* (2) & Organiz* (1) & Housekeeping (1) & Fix* regression (1) \\

 \midrule
\textit{Other branches} \\  \hline   \\
Refactor* (3669) & Chang* (2113) & Mov* (599) & Add* (519) & Fix* (427) \\
Remov* (342) & Creat* (253) & Simplif* (96) & Introduc* (93) & Renam* (89) \\
Replac* (71) & Split* (70) & Cleanup (68) & Improv* (64)  & Extract* (50) \\
Modif* (48) & Merg* (40) & Reduc* (33) & Extend* (26) & Rewrit* (23) \\
Get rid of (22) & Remov* unused (17) & Dependenc* (14) & Inlin* (13) & Organiz* (12) \\
Encapsulat* (9) & Restructur* (8) & Remov* redundant (8) & Enhanc* (7) & Code clean* (6) \\ 
Reformat* (5) & Rework* (4) & Reorder* (3) & Modulariz* (2) & Polish* (2) \\
Reorganiz* (2) & Fix regression (3) & Cosmetic chang* (2) & Customiz* (2) & Re-writ* (1) \\ 
Less code (1) & Chang* the name (1) & Code clarity (1) &  \\ \hline
\end{tabular}
\end{adjustbox}
\vspace{-.2cm}
\end{table} 

\subsection{\textcolor{black}{\RQtwo}}

\noindent\textbf{Motivation.} \textcolor{black}{Since there is no consensus on how
to formally document the act of refactoring code \cite{alomar2019can,alomar2021ESWA}, research question two identifies what textual patterns developers have used to describe their refactoring activities in `Refactor' branch. Identifying these patterns can reveal how developers communicate their intentions and needs, which is crucial to improving documentation and review processes.}

\noindent\textbf{Approach.} \textcolor{black}{We manually inspect Qt's subject and description to
identify refactoring documentation patterns in refactoring reviews from `Refactor' branch and refactoring reviews from other branches containing the keyword `refactor*' in the subject and description. These patterns are represented as a keyword or phrase that frequently occurs in refactoring reviews.}

\noindent\textbf{Results.} \textcolor{black}{Our in-depth inspection resulted in a list of  29 and 43 refactoring documentation patterns for refactoring reviews from `Refactor' branch and other branches, respectively, as shown in Table~\ref{Table:GeneralPatterns}. Our findings show that the names of refactoring operations (\textit{e.g.}, `mov*', `renam*', `extract*') occur in the top frequently occurring patterns, and these patterns are mainly linked to code elements at different levels of granularity such as classes, methods, and variables.  These specific terms are well-known software refactoring operations and indicate developers' knowledge of the catalog of refactoring operations. We also observe that the top-ranked refactoring operation-related keywords include `mov*', `renam*', and `extract*'. Moreover, we observe the occurrences of refactoring specific terms such as `cleanup', `get rid of', and `remov* unused'.}


\textcolor{black}{RQ$_2$ indicates that developers tend to use limited textual patterns to document their refactorings in the `Refactor' branch compared to refactoring reviews from other branches. These patterns can provide either (1) a generic description of problems developers encounter or (2) a specific refactoring operation name following Fowler's names \cite{Fowler:1999:RID:311424}. Although previous studies show that rename refactorings are a common type of refactoring, \textit{e.g.}, \cite{alomar2022exploratory}, we notice that `mov*' and `extract*' are also among the topmost documented refactorings in `Refactor' branch and other branches. This can be explained by the fact that developers tend to make many design improvement decisions, including remodularizing packages by moving classes, reducing class-level coupling, and increasing cohesion by moving methods. This information can provide valuable references for the practice of refactoring documentation. For example, whether refactoring-related reviews have relevant information is a critical indicator of the quality of refactoring reviews.}

\begin{boxK}
\textit{\textbf{Summary.} \textcolor{black}{Within the `Refactor’ branch, developers tend to use fewer keywords in their documentation. This could be attributed to their specialized knowledge and expertise in refactoring techniques, which might result in more precise and less confusing documentation for reviewers.} 
}
\end{boxK}






\subsection{\textcolor{black}{\RQthree}}
\begin{table}
\fontsize{6.8}{6.8}\selectfont
\begin{center}
\caption{Summary of refactoring patterns in `Refactor' branch, clustered by refactoring related categories.}
\label{Table:TopSpecificKeywords}
\begin{tabular}{lll}

\toprule
\textbf{Internal QA (\%)} &
\textbf{External QA  (\%)} &
\textbf{Code Smell  (\%)} \\

\midrule
\textit{`Refactor' branch} \\  \hline \\
    
  Dependency (90.32\%)    &  Compatibility (70\%) &  Dead Code (3.22\%)  \\ 
   Inheritance (3.22\%)     & Flexibility (20\%)     &    \\ 
   Abstraction (3.22\%)     & Performance (10\%)  &  \\ 
   Complexity (3.22\%)    &    &   \\ 

 \midrule
\textit{Other branches} \\  \hline   \\
Inheritance (30.43\%) & Accessibility (35.63\%) & Code duplication (79.16\%)\\
Dependency (30.43\%) & Performance (24.13\%) & Dead Code (10.41\%)  \\
Coupling (19.56\%) & Readability (18.39\%) & Switch Statement (4.16\%)\\
Abstraction (10.86\%) & Compatibility(6.89\%) & Antipattern (4.16\%) \\
Complexity (6.52\%) & Correctness (6.89\%) & Data Class (2.08\%) \\
Composition (2.17\%) & Modularity (2.29\%) & \\
& Stability (1.14\%) & \\
& Usability (1.14\%) & \\
& Flexibility (1.14\%) & \\
& Robustness (1.14\%) & \\
& Testability (1.14\%) & \\  \hline
\end{tabular} 
\end{center}
\vspace{-.6cm}
\end{table}

\noindent\textbf{Motivation.} \textcolor{black}{Various studies have explored the bad programming
practices that trigger refactoring and the potential quality
attributes that are optimized when restructuring the code.
In this research question, we investigate whether developers
explicitly mention the purpose of their refactoring activity, \eg
improving structural metrics to fix code smells. Knowing the quality attributes prioritized by developers helps to understand their focus areas and can guide the creation of better refactoring guidelines and tools.}

\noindent\textbf{Approach.} \textcolor{black}{After identifying refactoring documentation patterns, we categorize the patterns into main categories (similar to previous studies \cite{alomar2019can,alomar2021ESWA}): (1) internal quality attributes, (2) external quality attributes, and (3) code smells.}

\noindent\textbf{Results.} \textcolor{black}{Table \ref{Table:TopSpecificKeywords} provides the list of refactoring documentation patterns, ranked based on their frequency, which we identify in `Refactor' branch and other branches. We observe that developers mention key internal quality attributes (such as `inheritance', `complexity', etc.), a few external quality attributes (such as `compatability' and `performance'), and code smells (such as `dead code') that might impact code quality. To improve internal design, optimization of the structure of the system with respect to its dependency and inheritance appears to be the dominant focus, which is consistently mentioned in the review. Concerning external quality attributes, we observe the mention of refactorings to enhance nonfunctional attributes. Patterns such as `compatability', `flexibility', and `performance' represent the the main focus in `Refactor' branch, with 70\%, 20\%, and 10\%, respectively. Finally, for code smells, developers mentioned a few antipatterns such as `dead code'.}

\textcolor{black}{From RQ$_3$, we observe that developers in the `Refactor' branch tend to provide less explicit documentation of their intent compared to other branches. 
 This could suggest that the `Refactor' branch prioritizes higher code quality, allowing reviewers to grasp the context and purpose of changes more easily even without detailed documentation. For instance, developer discussed fixing design issues by putting common functionalities into a superclass to eliminate duplicate code, breaking up lengthier methods to make the code more readable, and avoiding nested complex data structure to reduce code complexity. Moreover, we observe that code smell is rarely documented in `Refactor' branch with only 3.22\%.  Similarly, developers tend to report few external quality attributes, focusing mainly on fixing \textit{compatability} of the code.}

\begin{boxK}
\textit{\textbf{Summary.} \textcolor{black}{Documentation of intent by developers is notably limited within the `Refactor' branch when compared to other branches. However, this observation may signify higher code quality within the `Refactor' branch, as changes are clearer and more understandable to reviewers without explicit documentation. This suggests that developers in the `Refactor' branch prioritize key quality attributes such as compatibility, performance, complexity, and inheritance.}}
\end{boxK}

\subsection{\RQfour}

\noindent\textbf{Motivation.} \textcolor{black}{We pose this research question to
develop a taxonomy of all refactoring contexts, where reviewers discuss refactoring in the `Refactor' branch. Analyzing discussion topics provides insights into the critical issues and considerations during refactoring reviews, highlighting areas for potential improvement.}

\noindent\textbf{Approach.} \textcolor{black}{To get a more qualitative sense, we manually inspect the Qt ecosystem using a thematic analysis technique \cite{cruzes2011recommended}, to study the topics that reviewers discuss when reviewing refactoring changes, so we understand the main reasons for which refactoring reviews does not take shorter compared to non-refactoring reviews or refactoring reviews from other branches.}

\begin{sidewaysfigure}[htbp]
\centering 
\includegraphics[width=\textwidth]{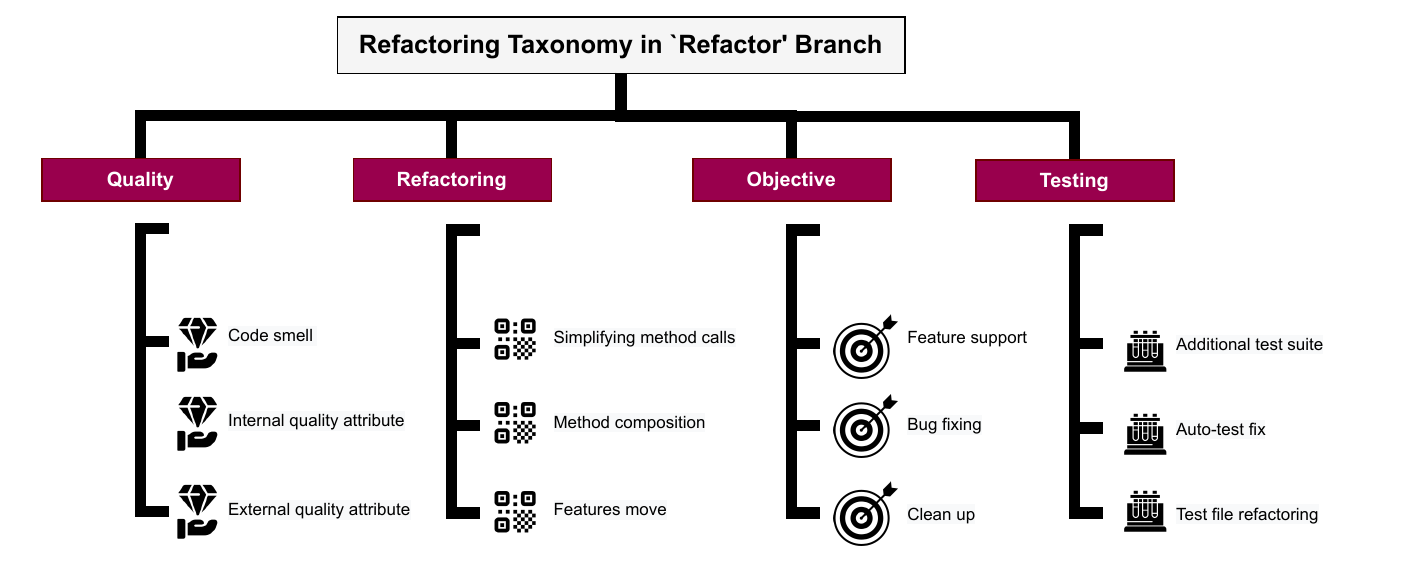}
\caption{\textcolor{black}{Refactoring review criteria from ‘Refactor’ branch in modern code review.}}
\label{fig:challenges-branch}
\end{sidewaysfigure}

\noindent\textbf{Results.} \textcolor{black}{Upon analyzing the review discussions, we create a comprehensive high-level categories of review criteria. Figure \ref{fig:challenges-branch} shows the proposed taxonomy of the criteria related to reviewing the refactored code. The taxonomy is composed of two layers: The top layer contains 4 categories that group activities with similar purposes, whereas the lower layer contains 12 subcategories that essentially provide a fine-grained categorization. These refactoring review criteria are centered on four main categories, 
  as shown in the figure: (1) quality, (2) refactoring, (3) objective, and (4) testing.  It is worth noting that our categorization is not mutually exclusive, meaning that a review can be associated with more than one category. An example of each category is provided in Table \ref{Table:example}. In the rest of this subsection, we provide a more in-depth analysis of these categories.}

  \noindent\textbf{Quality.} \textcolor{black}{The quality of design emerges as a crucial aspect of the refactoring review process. As per the submitted reviews, developers delve into optimizing \textit{internal} and \textit{external quality attributes} while striving to avoid \textit{code smells}. They offer recommendations on coding practices and suggest ways to enhance both internal and external quality attributes. This attention to detail is essential because developers may not always grasp the full scope of the software design. Additionally, developers often focus their refactoring efforts on classes and methods that undergo frequent changes. This pattern is evident in the reviews, where recurrent files are frequently mentioned. By repeatedly modifying the same code elements, developers become more intimately acquainted with the system, thereby enhancing their design decisions.}

  \noindent\textbf{Refactoring.} \textcolor{black}{This category gathers reviews with a focus on evaluating the correctness of the code transformation and checking whether or not the submitted changes lead to a safe and trustworthy refactoring. These refactoring reviews discuss refactoring operation-related responses such as \textit{simplifying method calls}, \textit{method composition}, and \textit{features move}. As developers often interleave refactoring with other tasks, developers mentioned that combining refactoring with other changes could potentially result in overshadowing errors, thus increasing the likelihood of introducing bugs.}

  \noindent\textbf{Objective.}  
\textcolor{black}{In this category, we have gathered cases where developers document the \textit{feature-related}, \textit{bug fix-related}, and  \textit{clean up-related} activities to better understand the rationale of the submitted code changes. This reveals how developers keep proposing areas of improvement pertaining to the perception and the rationale of the change. It appears that the clarity of the documented changes is of paramount importance in accepting the submitted refactoring changes. We realized that the clarity of the explanation of what is being changed and why affects review time and decision.}  

\noindent\textbf{Testing.} \textcolor{black}{Refactoring is intended to maintain the behavior of the software. Ideally, utilizing existing unit tests to confirm that the behavior remains unchanged should suffice. However, refactoring tasks may sometimes be interleaved with other activities, leading to potential alterations in the software's behavior. In such cases, existing unit tests may not capture these changes if they haven't been revalidated to reflect the newly introduced functionality. Upon analyzing these discussions, reviewers have proposed several recommendations. They suggest incorporating unit tests before initiating the refactoring process to instill greater confidence that the code remains intact. Additionally, they recommend adding test cases when refactored code results in decreased test coverage, such as when extracting new methods. Furthermore, when developers submit their changes for review, including the results of running the tests can enhance transparency and provide assurance to reviewers. To encompass these scenarios, we have identified the following sub-categories under the \textit{Testing} category: \textit{additional test suite}, \textit{auto-test fix}, and \textit{test file refactoring}.}

\textcolor{black}{
Our taxonomy builds on and extends existing refactoring taxonomies by focusing specifically on the unique characteristics of the `Refactor' branch. Unlike general refactoring taxonomies that encompass a wide range of refactoring activities across various contexts, our taxonomy is designed to capture specific practices and challenges and review dynamics within a dedicated refactoring branch. This taxonomy can be compared with existing ones in the literature, such as those proposed by Pantiuchina \etal \cite{pantiuchina2020developers,pantiuchina2021developers} and Paixao \etal \cite{paixao2020behind}, which classify refactorings based on reasons for refactoring rejection and rationale in the GitHub pull requests and Gerrit, respectively. For example, previous studies found that lack of clear goals and poorly documented proposals are the main causes of rejection after code review, while design improvement and test quality are key motivations. Also, they found that non-explicit or mixed intent refactorings tend to have high interactivity during the review, observations we shared when creating the taxonomy in the `Refactor' branch.}

\textcolor{black}{Unlike traditional taxonomies, ours is tailored to the unique context of the `Refactor' branch, reflecting its specific practices and priorities. By focusing on the `Refactor' branch, the taxonomy provides a more granular and context-specific understanding of refactoring practices. It offers insights into how branch isolation and focused collaboration can streamline the review process, reduce review time, and improve code quality. These findings are particularly valuable for organizations considering the adoption of dedicated refactoring branches as a strategy to enhance their development workflows. The outcome of the survey with a senior developer shows the existence
of these types of review criteria in the `Refactor' branch.}

\textcolor{black}{Furthermore, we conducted a manual analysis to identify the factors that contribute to successful reviews in the refactoring branch compared to other branches. These insights were drawn from developer documentation and our observations of best practices. The following best practices were identified:}

\noindent\textcolor{black}{\textbf{Focused Scope of Changes.} Reviews in the `Refactor' branch often involve changes that are narrowly focused on improving code structure without altering functionality. This clear separation may help reviewers quickly grasp the intent and impact of the changes. For example, in review ID 4058 (see Figure \ref{fig:example-casestudies}), the developer focused solely on renaming variables to improve code readability. The review had only two inline comments and was approved within 2.41, demonstrating the efficiency of focused changes.}

\noindent\textcolor{black}{\textbf{Clear and Concise Documentation.} Reviews in the `Refactor' branch often include clear, concise documentation explaining the purpose of the refactoring, the specific changes made, or the expected benefits. For instance, in review ID 1733 (see Figure \ref{fig:example-casestudies}), the developer provided a clear explanation of the purpose of refactoring, and a succinct summary of the changes, which led to a smooth review process with minimal comments.}

\noindent\textcolor{black}{\textbf{Consistency with Coding Standards.} Reviews in the `Refactor' branch often adhere to Gerrit Qt's established coding standards and guidelines, making it easier for reviewers to evaluate changes without debating style or format issues. For example, we noticed in review ID 4058 (see Figure \ref{fig:example-casestudies}) that the developer used Qt specific naming convention when naming classes.}


\begin{boxK}
\textit{\textbf{Summary.} \textcolor{black}{Discussions within the `Refactor' branch primarily revolve around topics related to code quality, objectives and testing. In particular, developers actively engage in conversations to ensure that the refactoring efforts adhere to the coding standards, discussions center on clarifying the objectives of the refactoring task, and there is a focus on rigorous testing practices to verify that the refactoring changes do not introduce new issues and that the improved code performs as expected.}
}
\end{boxK}

\begin{table*}[htbp]
  \centering
	 \caption{A taxonomy of the refactoring review criteria from `Refactor' branch in modern code review.} 
	 \label{Table:example}
\begin{sideways}
\begin{adjustbox}{width=1.1\textwidth,center}
\begin{tabular}{llLllll}\hline
\toprule
\bfseries Category & \bfseries Sub-category & \bfseries Example (Excerpts from a related refactoring review) \\
\midrule
\multirow{5}{*} {\textbf{Quality}} & \cellcolor{gray!30}Code smell & \cellcolor{gray!30} \say{\textit{Qt Creator: remove dead patching code  We are using a VCS. No need to keep dead code.}} \\ 
& Internal quality attribute & \say{\textit{Checking in text control and editor classes what of those related attributes have exactly changed before calling the update increases code complexity unnecessarily for a little benefit.}}\\
& \cellcolor{gray!30}External quality attribute & \cellcolor{gray!30} \say{\textit{This is simpler than the existing texture cache in QtOpenGL, as it only serves the GL paint engine. There's one per context group, to simplify the design and to prevent performance degradations.}}\\
\hline
\multirow{5}{*} {\textbf{Refactoring}} & Simplifying method calls& \say{\textit{Refactor QMouseEvent to contain the position inside the window. Rename the default accessors for positions to localPos, windowPos
and screenPos, to be explicit about their use.}} \\ 
& \cellcolor{gray!30}Method composition & \cellcolor{gray!30} \say{\textit{Adding 10s delay after extracting 7z.}}\\
& Feature move  & \say{\textit{Add QtUiTools and uilib  This code used to live in qtbase. It doesn't belong there however and since there are no dependencies in qtbase left that require it move it here.  This significantly simplifies the build system with regards to the code in uilib.}}\\
 \hline
\multirow{7}{*} {\textbf{Objective}} & \cellcolor{gray!30}Feature support & \cellcolor{gray!30}\say{\textit{Restore feature compatibility with QPrinter in QTextDocument::print  Add a margin method to QPagedPaintDevice.}} \\ 
&Bug fixing & \say{\textit{fix foundImportantUpdate and rename it to essential   - there was bug that unselected updates are removed in case there is an important update   - to avoid the error with old installation we are renaming Important tag to Essential which means the same but only works with this fix.}}\\ 
& \cellcolor{gray!30}Clean up & \cellcolor{gray!30} \say{\textit{Refactor the input framework  Results of the ongoing workshop in Oslo: QInputPanel will be the application facing interface for controlling the input context as well as querying things like the position of the virtual keyboard. QInputContext is significantly cleaned up and only there as a compatibility API for existing code.}}\\ 
\hline
\multirow{4}{*} {\textbf{Testing}} &Additional test suite& \say{\textit{Added new QOpenGLPaintDevice test case in tst\_QOpenGL.}} \\
& \cellcolor{gray!30}Auto-test fix & \cellcolor{gray!30} \say{\textit{Fixed auto-test failure in tst\_QOpenGL.  QOpenGLFramebufferObject::height() was returning the width.}}\\
&Test file refactoring & \say{\textit{Move tests to more logical positions.  Before, all the auto tests were jumbled together in a huge mess in tests/auto, now they are organized after which module/submodule/class they belong to.  I have also started separating out unit tests from integration tests.}}\\

\bottomrule
\end{tabular}
\end{adjustbox}
\end{sideways}
\vspace{-.3cm}
\end{table*}
\section{Implications}
\label{Section:Implications}

\subsection{Implications for Practitioners}
\textcolor{black}{\textbf{Establishing continuous improvement culture for refactoring-related reviews.} Our RQ$_1$ findings show that developers are more inclined to accept refactoring changes quickly in the refactoring branch compared to the refactoring changes in other branches or to the non-refactoring changes. To emphasize the value of refactoring changes, managers can cultivate a culture by prioritizing continuous improvement and investing in refactoring. Therefore, establishing guidelines for submitting refactoring changes for review can be beneficial. For instance, refactoring branches should involve smaller, focused changes aimed at improving specific areas of the code. Moreover, these branches should emphasize the improvement of software quality, aligning with the shared goal among developers.  It is essential for refactoring branches to clearly document the primary goals of the proposed changes. Additionally, thorough code reviews and testing should be conducted to ensure adherence to best practices and to prevent the introduction of regressions or breaking existing functionality. \textcolor{black}{In summary, the recommended checklists are as follows:}}


\color{black}
\section*{Guidelines for Submitting Refactoring Changes to the `Refactor' Branch}

\begin{itemize}[label=$\square$]
    \item \textbf{Nature of Changes:}
    \begin{itemize}[label=$\checkmark$]
        \item Are the changes primarily focused on improving the structure, readability, or maintainability of the code without altering its external behavior?
        \item Do the changes involve renaming variables, methods, or classes to improve clarity?
        \item Are the changes aimed at reducing technical debt by simplifying complex code or removing redundancies?
    \end{itemize}

    \item \textbf{Scope of Changes:}
    \begin{itemize}[label=$\checkmark$]
        \item Do the changes impact multiple files or modules, indicating a broader structural improvement?
        \item Are the changes part of a planned refactoring effort as outlined in the project's roadmap or guidelines?
    \end{itemize}

    \item \textbf{Documentation and Communication:}
    \begin{itemize}[label=$\checkmark$]
        \item Have you documented the intent, instruction, impact, and scope of the refactoring changes in the commit message or accompanying documentation?
        \item Have you communicated with the team about the planned refactoring and received approval to proceed with the changes?
    \end{itemize}

    \item \textbf{Testing and Validation:}
    \begin{itemize}[label=$\checkmark$]
        \item Have you ensured that all existing tests pass after the refactoring changes?
        \item Have you added or updated tests to cover the refactored code, ensuring no functionality is broken?
    \end{itemize}

    \item \textbf{Review Process:}
    \begin{itemize}[label=$\checkmark$]
        \item Have you considered the feedback from previous refactoring reviews and incorporated best practices into your changes?
    \end{itemize}
\end{itemize}

\section*{Guidelines for Submitting Changes to the Main Branch}

\begin{itemize}[label=$\square$]
    \item \textbf{Functional Changes:}
    \begin{itemize}[label=$\checkmark$]
        \item Are the changes introducing new features, fixing bugs, or modifying the external behavior of the application?
        \item Do the changes involve implementing new functionality or altering existing functionality to meet new requirements?
    \end{itemize}

    \item \textbf{Isolated Updates:}
    \begin{itemize}[label=$\checkmark$]
        \item Are the changes isolated to a single file or a small set of related files?
        \item Do the changes address a specific issue or feature request without requiring broader structural modifications?
    \end{itemize}
\end{itemize}
\color{black}
\textcolor{black}{By following these checklists, developers can make informed decisions about where to submit their changes, ensuring that the refactoring branch is used appropriately for structural improvements while functional changes are directed to the main branch. We believe this approach can enhance the clarity and effectiveness of the development process, benefiting both individual developers and the overall project. It is worth noting that this checklist has been validated by an external developer for its relevance and completeness. However, we plan to apply this checklist in practice to further assess its applicability and impact in real-world scenarios.}

\textbf{Establishing guidelines for refactoring-related reviews.} Our taxonomy shows that reviewing refactoring goes beyond improving the code structure. To improve the practice of reviewing refactored code, and contribute to the quality of reviewing code in general, managers can collaboratively work with developers to establish customized guidelines for reviewing refactoring changes which could establish beneficial and long-lasting habits or themes to accelerate the process of reviewing refactoring. Additionally, since our RQ$_4$ findings show that testing is one of the topics discussed by developers when reviewing refactoring changes, it is recommended to utilize continuous integration to keep the testing suite in sync with the code base during and after refactoring. 



\subsection{Implications for Researchers}
\textbf{Understanding how refactoring changes in the ‘Refactor’ branch tend to be reviewed.} From RQ$_1$, we observe that they refactoring changes in the ‘Refactor’ branch are completed in a shorter timeframe compared to changes from other branches and
non-refactoring changes. Researchers should further investigate the underlying reasons why refactoring changes in the `Refactor' branch tend to be reviewed more efficiently compared to changes from other branches. Understanding the factors contributing to this efficiency, such as branch isolation, developer expertise, or collaboration dynamics, can provide valuable information on optimizing code review processes and enhancing software quality in development environments where refactoring is a common practice. Furthermore, exploring the impact of streamlined refactoring review practices on code quality, developer productivity, and overall project success could offer practical guidance for software development teams aiming to improve their refactoring workflows.

\textbf{Supporting for the refactoring of non-source code artifacts.} From RQ$_4$, we discover that refactoring operations are not limited to source code files. Artifacts such as databases and log files are also susceptible to refactoring. Similarly, we also observed discussions about refactoring test files. While it can be argued that test suites are source code files,  recent studies \cite{pantiuchina2020developers,alomar2021ESWA} show that the types of refactoring operations applied to test files are frequently different from those applied to production files. Hence, future research on refactoring is encouraged to introduce refactoring mechanisms and techniques exclusively geared to refactoring non source code artifact types and test suites.


\subsection{Implications for Tool Builders}
\textbf{Developing next generation refactoring-related code review tools.}  Finding that reviewing refactoring changes from other branches takes longer than non-refactoring changes reaffirms the necessity of developing accurate and efficient tools and techniques that can assist developers in the review process in the presence of refactorings. Refactoring toolset should be treated in the same way as CI/CD tool set and integrated into the tool-chain. 
 Researchers could use our findings with other empirical investigations of refactoring to define, validate, and develop a scheme to build automated assistance for reviewing refactoring considering the refactoring review criteria as review code becomes an easier process if the code review dashboard augmented with the factors to offer suggestions to better document the review. 
  
 \textcolor{black}{Furthermore, to accelerate the code review process and limit having a back-and-forth discussion for clarity on the problem faced by the developer, tool builders can develop \textit{bots} for the integration, testing, and management categories. Additionally, it would be interesting to use a popular and widely adopted quality framework, \eg  Quality Gate of SonarQube \cite{olivier2013sonar}, as part of the quality verification process by embedding its results in the code review. This might facilitate convincing the reviewer about the impact and the correctness of the performed refactoring.}

\section{Threats To Validity}
\label{Section:Threats}

In this section, we describe potential threats to validity of our research method, and the actions we took to mitigate them.

\textbf{Internal Validity.} Concerning the identification of refactoring-related code review, we select reviews from a dedicated `Refactor' branch.  As for the other group of refactoring reviews, we analyze reviews with the keyword `\textit{refactor*}' in their title and description. Such selection criteria may have resulted in missing refactoring-related reviews, and there is the possibility that we may have excluded synonymous terms/phrases. However, even though this approach reduces the number of reviews in our dataset, it also decreases the false positiveness of our selection. 
 While our data collection may result in missing some reviews, our approach ensures that we analyze reviews that are explicitly geared toward refactoring. In other words, these are reviews where developers were explicitly documenting a refactoring action and they wanted it to be reviewed.  Additionally, after performing the manual inspection on review discussions, we realized that refactoring is heavily emphasized in discussions that start with a title or a description containing the keyword `\textit{refactor*}'. Yet, this does not prevent other discussions from bringing refactoring into the picture, and these will be missed by our selection (\ie false negatives). Hence, we excluded potential refactoring synonymous terms/phrases when selecting non-refactoring reviews. 
  We opted for such picky selection to only consider discussions when code authors explicitly wanted their refactored code to be reviewed, and so reviewers eventually propose a refactoring-aware feedback, which is what we are aiming for in this study. 
  Therefore, it is interesting to consider scenarios where reviewers have raised concerns about refactoring a code change that was not intended to be associated with the `Refactor' branch. Since refactoring can easily be interleaved with other functional changes, it would be interesting to extract scenarios where reviewers thought it was misused. The study can also help developers better understand not only how to refactor their code, but also how to document it properly for easier review.

Furthermore, we focus on the code review activity that is reported by the tool-based code review process, \ie Gerrit, of the systems studied due to the fact that other communication media (\eg in-person discussion \cite{beller2014modern}, a group IRC \cite{shihab2009studying}, or a mailing list \cite{rigby2011understanding}) do not have explicit links to code changes and recovering these links is a daunting task \cite{thongtanunam2016revisiting,bacchelli2010linking}.

\textbf{Construct Validity.} About the representativeness and the correctness of our refactoring review criteria, we derive these criteria from a manual analysis of refactoring-related reviews. This approach may not cover the whole spectrum of all the review criteria done with refactoring in mind. 
 Additionally, to avoid personal bias during the manual analysis, each step in the manual analysis was conducted twice, and the results were always cross-validated. Another potential threat
to validity relates to refactoring reviews. Since refactorings could interleave with other changes \cite{murphy2012we} (\ie developers performed changes together with refactorings), we cannot claim that the selected refactoring reviews are exclusively related to refactoring. However, during our qualitative analysis, we identified this activity as one of the challenges that contributed to slowing down the review process from other branches.

\textbf{External Validity.} We focus our study on one open-source system due to the low number of systems that satisfied our eligibility criteria (see Section \ref{Section:Methodology}). Therefore, our results may not be generalized to all other open-source systems or commercially developed projects. However, the goal of this paper is not to build a theory that applies to all systems, but rather to show that refactoring can have an impact on code review process. Another potential threat relates to the proposed taxonomy. Our taxonomy may not generalize to other open source or commercial projects since the refactoring review criteria may be different for another set of projects (\eg outside the Qt community). Consequently, we cannot claim that the results of refactoring review criteria (see Figure \ref{fig:challenges-branch}) can be generalized to other software systems where the need for improving the design might be less important. 

\textbf{Conclusion Validity.} To compare the two groups of code review requests, we used appropriate statistical procedures with \textit{p}-value and effect size measures to test the significance of the differences and their magnitude.  A statistical test was implemented to measure the significance of the observed differences between group values. This test makes no assumption that the data are normally distributed. Also, it assumes the independence of the groups under comparison. We cannot verify whether code review requests are completely independent, as some can be re-opened, or one large code change can be treated using several requests. To mitigate this, we verified all the reviews we sampled for the test.

\section{Conclusion}
\label{Section:Conclusion}

Understanding the practice of refactoring code review holds significant importance for both the research community and industry. Despite the widespread adoption of modern code review practices in open-source and industrial projects, the correlation between code review and refactoring practices in the `Refactor' branch remains largely unexplored. In our study, we conducted a comprehensive quantitative and qualitative analysis to investigate the review criteria discussed by the developers during the review of refactorings. Our findings highlight several key points: refactoring changes in the `Refactor' branch are completed in a shorter timeframe compared to changes from other branches and non-refactoring changes; documentation of developer intent within the `Refactor' branch is limited in comparison to other branches; and developers rely on a specific set of criteria to guide their decisions regarding the acceptance or rejection of submitted refactoring changes.

For future work, we plan on conducting a structured survey with software developers from both open-source and industry. The survey will explore their general and specific review criteria when performing refactoring activities in code review. This survey will complement and validate our current study to provide the software engineering community with a more comprehensive view of refactoring practices in the context of modern code review. Another interesting research direction is to link refactoring-related reviews to refactoring detection tools such as Refactoring Miner \cite{tsantalis2018accurate} or RefDiff \cite{silva2017refdiff} to better understand the impact of these reviews on refactoring types specifically.

\noindent{\textbf{Declaration of generative AI and AI-assisted technologies in the writing process.}}

During the preparation of this work, the author used the ChatGPT Web interface to improve the language and readability of some sections such as the checklists. After using this tool, the author reviewed and edited the content as needed and takes full responsibility for the content of the publication.

\bibliography{references}

\end{document}